\documentclass[]{interact}

\usepackage{epstopdf}
\usepackage[caption=false]{subfig} 

\usepackage[natbibapa,nodoi]{apacite} 
\theoremstyle{plain} 

\theoremstyle{definition}

\theoremstyle{remark}

\usepackage{algorithm}   
\usepackage{algorithmic} 
\usepackage{multirow}     
\usepackage{colortbl}       
\usepackage{url}
\usepackage{setspace}
\usepackage{pdflscape} 

\let\Algorithm\algorithm
\renewcommand\algorithm[1][]{\Algorithm[#1]\setstretch{1.2}}

\definecolor{kugray5}{RGB}{224,224,224}
\definecolor{navyblue}{rgb}{0.0, 0.0, 0.5}
\definecolor{blue2}{rgb}{0.16, 0.32, 0.75}
\definecolor{ao}{rgb}{0.0, 0.5, 0.0}
\definecolor{red2}{rgb}{0.7, 0.11, 0.11}

\usepackage{array}
\usepackage{xcolor}
\makeatletter
\newcommand*{\@rowstyle}{}
\newcommand*{\rowstyle}[1]{
  \gdef\@rowstyle{#1}%
  \@rowstyle\ignorespaces%
}
\newcolumntype{=}{
  >{\gdef\@rowstyle{}}%
}
\newcolumntype{+}{
  >{\@rowstyle}%
}
\makeatother

\begin{document}


\title{Adaptive Partitioning Design and Analysis for Emulation of a Complex Computer Code}

\author{
\name{Sonja Surjanovic and William J. Welch\thanks{Corresponding author: William J. Welch. Email: will@stat.ubc.ca}}
\affil{Department of Statistics, The University of British Columbia, Vancouver, BC, V6T 1Z4, Canada}
}

\maketitle

\begin{abstract}
Computer models are used as replacements for physical experiments in a large variety of applications. Nevertheless, direct use of the computer model for the ultimate scientific objective is often limited by the complexity and cost of the model. Historically, Gaussian process regression has proven to be the almost ubiquitous choice for a fast statistical emulator for such a computer model, due to its flexible form and analytical expressions for measures of predictive uncertainty. However, even this statistical emulator can be computationally intractable for large designs, due to computing time increasing with the cube of the design size. Multiple methods have been proposed for addressing this problem. We discuss several of them, and compare their predictive and computational performance in several scenarios. \\[1ex]
We then propose solving this problem using an adaptive partitioning emulator (APE). The new approach is motivated by the idea that most computer models are only complex in particular regions of the input space. By taking a data-adaptive approach to the development of a design, and choosing to partition the space in the regions of highest variability, we obtain a higher density of points in these regions and hence accurate prediction.
\end{abstract}






\section{Introduction}
\label{sec:introduction}

Computer models are used as replacements for physical experiments in a wide variety of applications. Nevertheless, the number of evaluations of the computer model is often somewhat limited due to the complexity and cost of the model. As a result, the computer model is often replaced by a faster statistical emulator to model the input-output relationship.

For decades, Gaussian processes (GPs) have been the almost ubiquitous choice for the statistical emulator \citep[e.g.][]{Sacksetal1989}. However, the GP itself is computationally expensive for large designs, since the matrix calculations involved in evaluating the likelihood become intractable. 

We begin by presenting an overview of the methods that have been developed for overcoming the computational complexity, and compare their performance in several settings. We then propose a new method for solving this 
problem, dividing the input space during both design and analysis
in an adaptive partitioning emulator (APE), 
and compare its performance. We see that it is competitive with the existing methodology, and has a much lower computational complexity than standard GP fitting.




\subsection{The GP model}
\label{subsec:thegpmodel}

The most commonly used strategy for modelling the output of a deterministic computer experiment is Gaussian process regression \citep[e.g.][]{Sacksetal1989, Currinetal1991, OHagan1992}. In this model, the $d$-dimensional inputs are denoted by $\bm{x} = (x_1, \ldots, x_d)$, and the output (or response) is treated as the realization of a random function $Y(\cdot)$ that has the density of a Gaussian process with mean $\mu(\cdot)$, 
variance $\sigma^2$, and positive-definite correlation function $\rho(\cdot, \cdot)$.

The correlation between the responses at any two points, $\rho(\bm{x}, \bm{x}') = \textnormal{Corr}[Y(\bm{x}), Y(\bm{x}')]$, is assumed to be a function of the locations $\bm{x}$ and $\bm{x}'$ alone, and is chosen such that points closer together are more correlated. In many cases, a further assumption is made that the correlation function is \emph{separable}, meaning that it can be written as a product of one-dimensional correlation functions. There are many valid separable correlation functions to choose from. For instance, the Mat\'ern correlation function with smoothness parameter $\kappa = 5/2$ is
\begin{equation}
\rho(\bm{x}, \bm{x}') = \prod_{j=1}^d \left(1 + \frac{\sqrt{5} \big| x_j - x'_j \big|}{\theta_j} + \frac{5 \big| x_j - x'_j \big|^2}{3 \theta_j^2} \right) \times \textnormal{exp}\left( -\frac{\sqrt{5} \big| x_j - x'_j \big|}{\theta_j} \right).
\nonumber
\end{equation}
The correlation function parameters are denoted by $\bm{\theta} = \left[ \theta_1, \hdots, \theta_d \right]^\top$.

We also generally assume a linear trend term for the mean:
\begin{equation}
\mu(\bm{x}) = \sum_{k=1}^p \beta_k f_k(\bm{x}), \\[-1ex]
\label{eq:mu}
\end{equation}
for some set of coefficients $\bm{\beta} = \left[ \beta_1, \hdots, \beta_p \right]^\top$ and a set of covariate functions $f_1(\cdot), \hdots, f_p(\cdot)$. A stationary mean is commonly assumed, implying that $p = 1$, $f_1(\bm{x}) = 1$ and thus $\mu(\bm{x}) = \beta_1$ for all $\bm{x}$.

\subsubsection{Costs of parameter estimation} 
\label{subsubsec:costsofparameterestimation}

The parameters to be estimated in the Gaussian process regression model are $\bm{\beta}$, $\sigma^2$ and $\bm{\theta}$. Data are obtained on a set of $n$ design points $\mathcal{X} = \{ \bm{x}^{(1)}, \hdots, \bm{x}^{(n)} \}$, and the parameters are estimated using maximum likelihood, as in \cite{Jonesetal1998}. The log-likelihood function for the data is
\begin{equation}
\ell(\bm{\beta}, \sigma^2, \bm{\theta}) = \frac{n}{2} \text{log}(2 \pi \sigma^2) - \frac{1}{2} \text{log}(\text{det}(\bm{R})) - \frac{1}{2 \sigma^2} (\bm{y} - \bm{F} \bm{\beta})^\top \bm{R}^{-1} (\bm{y} - \bm{F} \bm{\beta}),
\label{eq:loglikelihood}
\end{equation}
where $\text{det}(\cdot)$ denotes the determinant of a matrix, $\bm{R}$ is the $n \times n$ correlation matrix for the data $\bm{y} = \left[ y(\bm{x}^{(1)}), \hdots, y(\bm{x}^{(n)}) \right]^\top$, and $\bm{F}$ is the $n \times p$ matrix whose $(i, k)^{\text{th}}$ entry is $f_k(\bm{x}^{(i)})$, the value of the $k^{\text{th}}$ covariate function at the $i^{\text{th}}$ observed location.

Assuming that the correlation parameters $\bm{\theta}$ are known, the maximum likelihood estimates of $\bm{\beta}$ and $\sigma^2$ can be written as
\begin{align}
\hat{\bm{\beta}}(\bm{\theta}) &= \left( \bm{F}^\top \bm{R}^{-1} \bm{F} \right)^{-1} \bm{F}^\top \bm{R}^{-1} \bm{y}, \text{ and} \label{eq:betahat}\\
\hat{\sigma}^2(\bm{\theta}) &= \frac{1}{n} (\bm{y} - \bm{F} \hat{\bm{\beta}}(\bm{\theta}))^\top \bm{R}^{-1} (\bm{y} - \bm{F} \hat{\bm{\beta}}(\bm{\theta})). \label{eq:sigma2hat}
\end{align}
Substituting these maximum likelihood estimates into \eqref{eq:loglikelihood} yields the \emph{concentrated log-likelihood function}, which is a function of only $\bm{\theta}$. This function is then numerically optimized to obtain the estimate $\hat{\bm{\theta}}$, which is then substituted back into \eqref{eq:betahat} and \eqref{eq:sigma2hat} to give the estimates $\hat{\bm{\beta}}$ and $\hat{\sigma}^2$.

This parameter estimation procedure can be very costly. Specifically, at every step of the numerical optimization, the inverse of the correlation matrix $\bm{R}$ must be calculated. For large design sizes $n$, this can be a very time-consuming procedure, because the matrix inversion is $\mathcal{O}(n^3)$, and it has to be repeated many times in the numerical search for the maximum likelihood estimates (similarly for Bayesian MCMC). Computing $\text{det}(\bm{R})$ within \eqref{eq:loglikelihood} is also $\mathcal{O}(n^3)$, although a single $\mathcal{O}(n^3)$ Cholesky factorization can give both the inverse and the determinant.

\subsubsection{Costs of prediction} 
\label{subsubsec:costsofprediction}

For the Gaussian process model defined above, the predictor that minimizes the Mean Squared Prediction Error (MSPE) is, for any location $\bm{x}_0$,
\begin{equation}
\hat{y}(\bm{x}_0) = \mu(\bm{x}_0) + \bm{r}^\top(\bm{x}_0) \bm{w}, \\
\label{eq:yhat}
\end{equation}
where $\bm{w}$ is an $n$-vector of weights, and $\bm{r}(\bm{x}_0) = [ \rho(\bm{x}^{(i)}, \bm{x}_0) ]_{i=1,\hdots,n}$ is the vector of correlations between $Y(\bm{x}_0)$ and each of the response values in the design. The weights are given by
\begin{equation}
\bm{w} = \bm{R}^{-1} (\bm{y} - \bm{F} \bm{\beta}). \\
\label{eq:w}
\end{equation}
Thus, even if the Gaussian process parameters $\{ \bm{\beta}, \sigma^2, \bm{\theta} \}$ are known (and thus do not need to be estimated), obtaining predictions also requires one inversion of the correlation matrix.




\subsection{Solving the large-dataset problem}
\label{subsec:solvingthelargedatasetproblem}

The GP regression model is a flexible statistical emulator for complex computer models: those with moderately high dimensionality, high degrees of non-linearity, or many higher-order interactions. However, a complex computer model may require thousands of runs before the GP model can accurately represent its behaviour.

As seen above, the matrix calculations involved in fitting and prediction using a GP model with moderately large design sizes can quickly become intractable. In particular, many likelihood evaluations are required for maximum likelihood or Bayesian MCMC estimation of the GP parameters, while each likelihood evaluation is quite expensive.


In Section~\ref{sec:gaussianprocessregressionwithlargedatasets}, we present an overview of the existing methods that have been proposed for easing the computational burden of fitting the GP regression model for large designs. In Section~\ref{sec:adaptivepartitioningemulator}, the new APE method is proposed. Section~\ref{sec:results} compares the performance of APE to  previous methods, using several test functions. Finally, a discussion is given in Section~\ref{sec:discussion}.





\section{Gaussian process regression with large datasets: Has the problem been solved?}
\label{sec:gaussianprocessregressionwithlargedatasets}

For complex computer models, moderate design sizes are required for the GP regression to accurately capture the behaviour of the function. However, as we have seen in the previous section, the GP model becomes intractable for such large designs. Some of the methods that have been proposed for solving this issue are described below.

The methods outlined here can be grouped into two categories: design-based methods and localization-based methods. In design-based methods, a specific type of design is paired with the standard GP analysis. Generally, the design is chosen in such a way as to ease the computational burden of the analysis calculations. In localization-based methods, the computational problem is addressed by breaking the analysis down into smaller fits, using one subset of the design at any given time. The subsets are usually chosen such that they contain points in close proximity to each other, or in close proximity to a desired prediction location.

The methods are outlined below, along with several details pertaining to their implementation in {\sf R} \citep{RCoreTeam2017} or MATLAB \citep{MATLAB2017}. Further detail on the calculations involved in each method can be found in the respective papers.




\subsection{Sparse Grid Designs}
\label{subsec:sparsegriddesigns}

The Sparse Grid Design \citep[SGD,][]{Plumlee2014} is a design-based approach. The main intuition here is to exploit the structure of a grid-like design to partition the matrix calculations in \S~\ref{subsubsec:costsofparameterestimation} into smaller steps.

In particular, a full grid is defined as the Cartesian product $\mathcal{X} = \mathcal{X}_1 \times \mathcal{X}_2 \times \hdots \times \mathcal{X}_d$, 
where each $\mathcal{X}_j = \{ x_j^{(1)}, \hdots, x_{j}^{(n_j)} \}$ (for $j = 1, \hdots, d$) is the one-dimensional \emph{component design} in the $j^{\text{th}}$ dimension, containing $n_j$ points. 
Then, it can be shown that the $n \times n$ correlation matrix for the full grid can be written as $\bm{R} = \otimes_{j=1}^d \bm{R}_j$, and thus $\bm{R}^{-1} = \otimes_{j=1}^d \bm{R}_j^{-1}$, where $\bm{R}_j$ is the $n_j \times n_j$ correlation matrix for $\mathcal{X}_j$ and $\otimes$ denotes a Kronecker product \citep{Plumlee2014}.

\begin{figure}[t!]
\centering
\subfloat{%
\resizebox*{4.55cm}{!}{\includegraphics{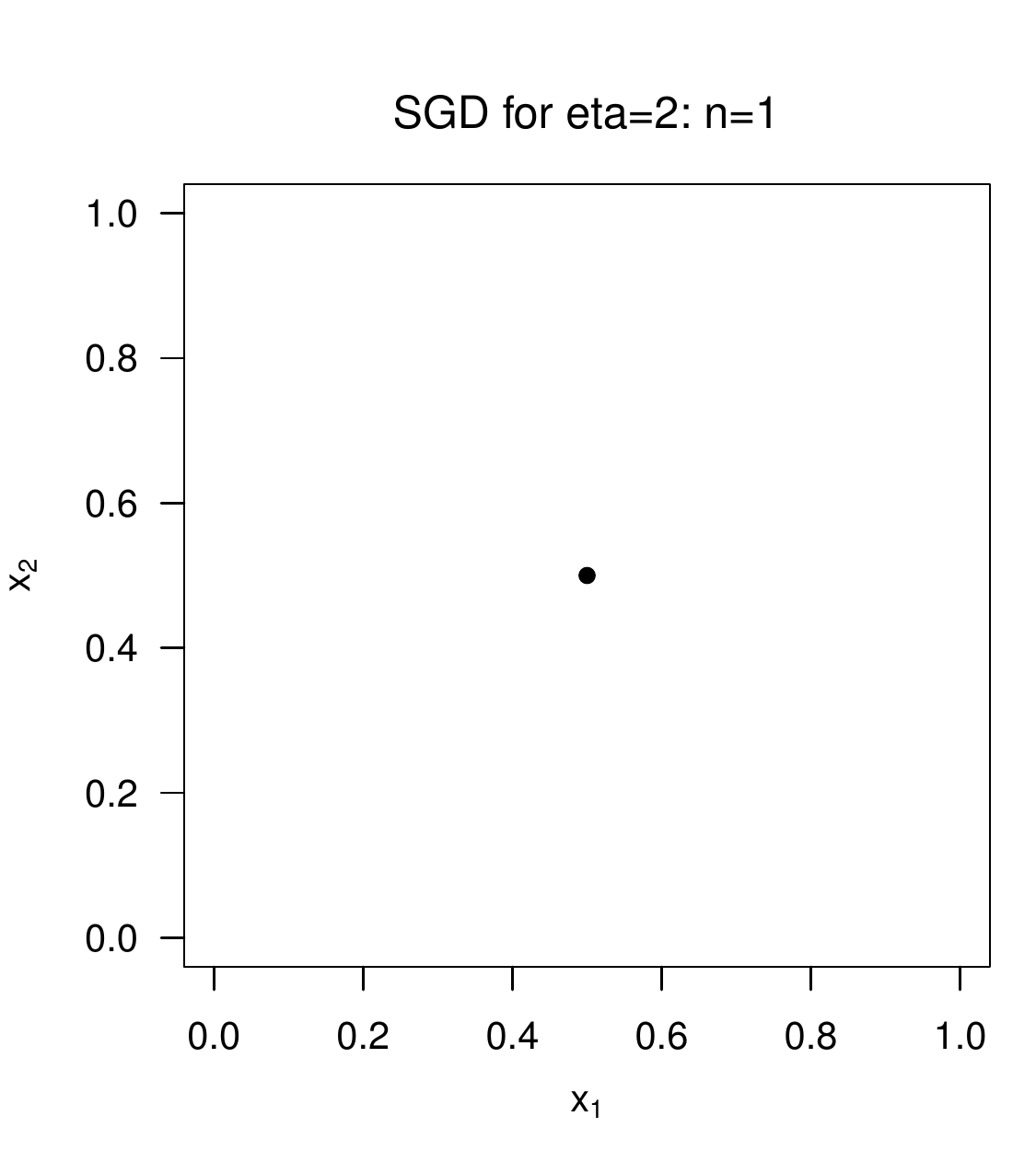}}}\hspace{5pt}\vspace{-3ex}
\subfloat{%
\resizebox*{4.55cm}{!}{\includegraphics{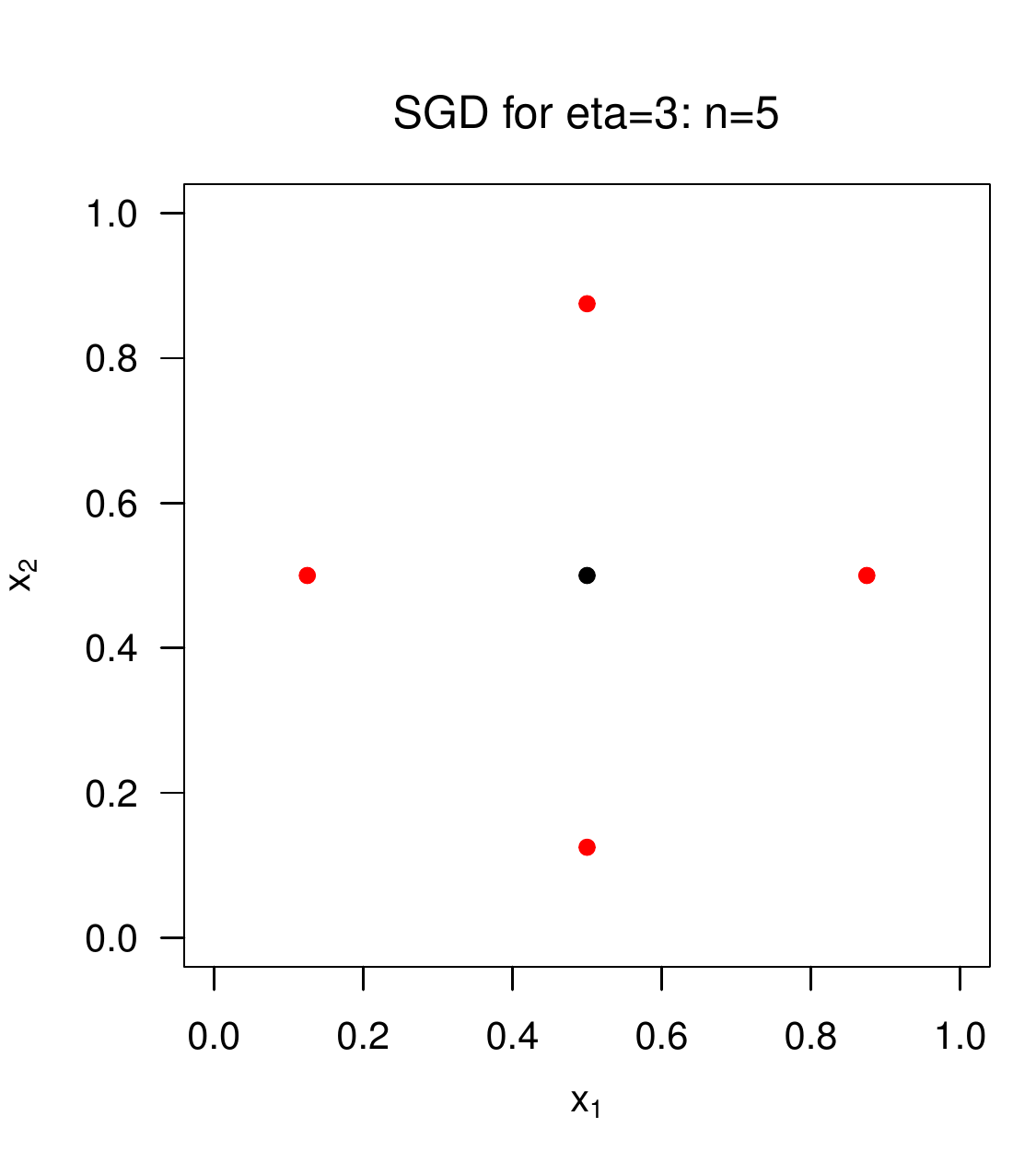}}}\hspace{5pt}
\subfloat{%
\resizebox*{4.55cm}{!}{\includegraphics{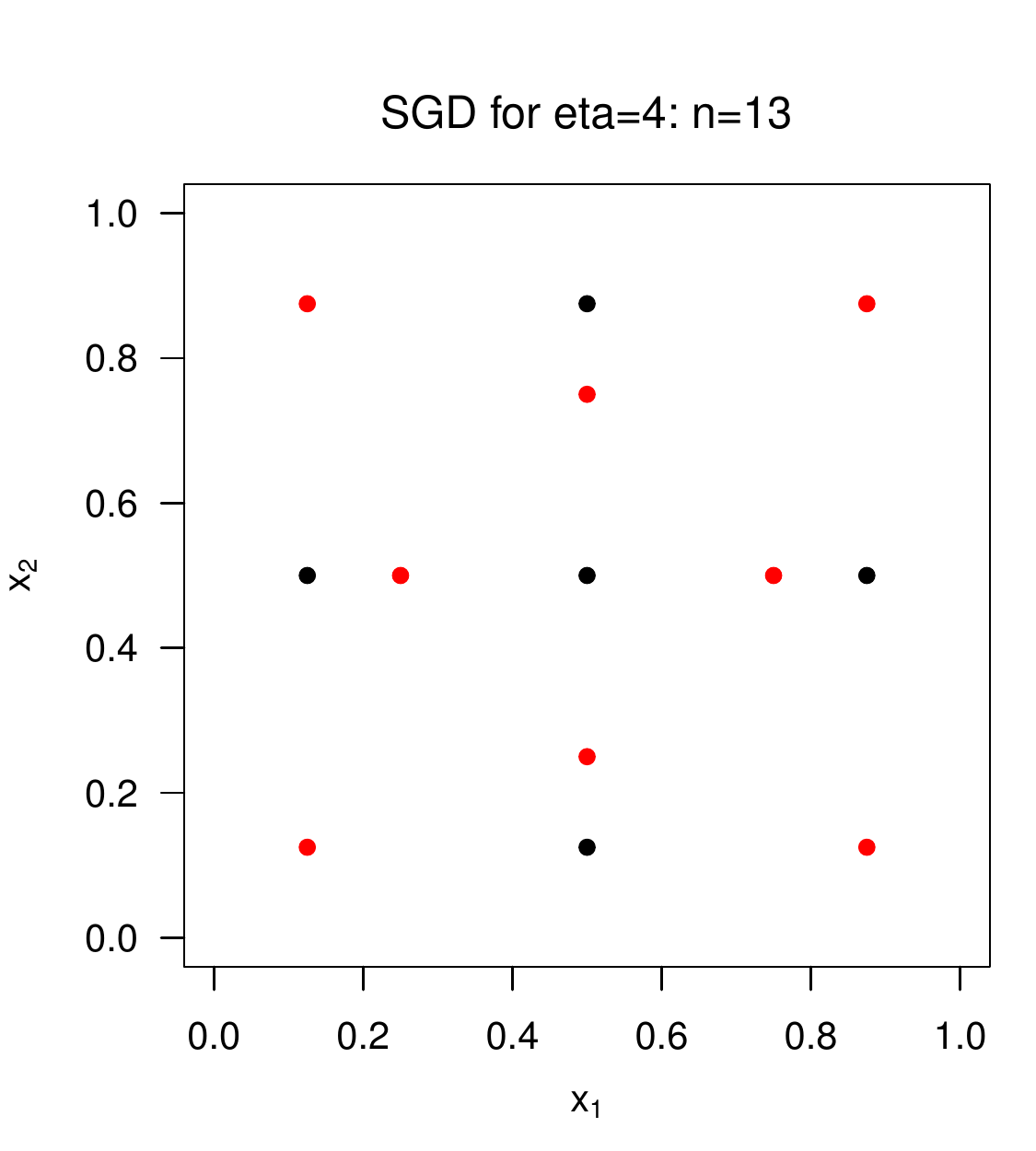}}}\hspace{5pt}
\subfloat{%
\resizebox*{4.55cm}{!}{\includegraphics{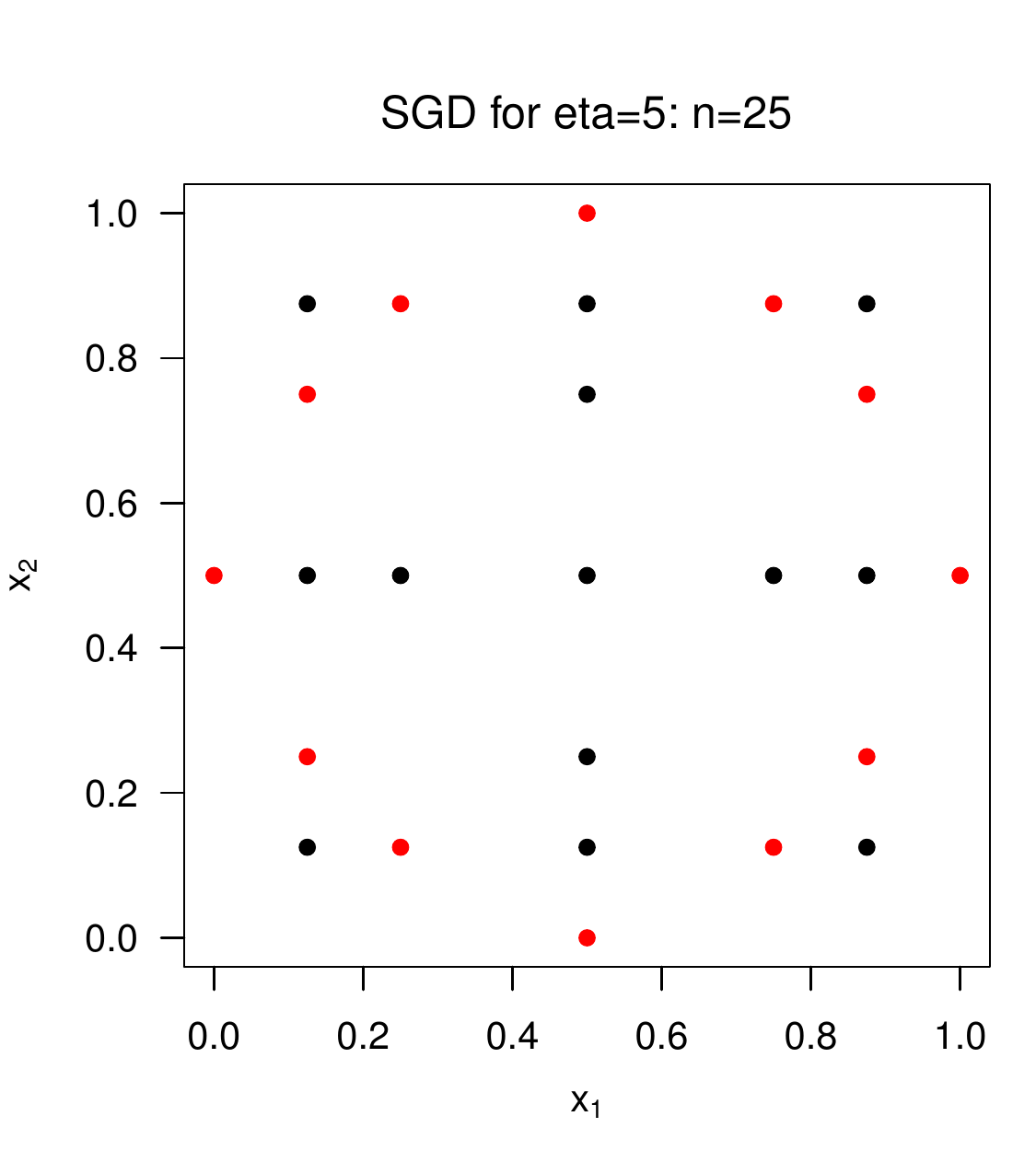}}}\hspace{5pt}
\subfloat{%
\resizebox*{4.55cm}{!}{\includegraphics{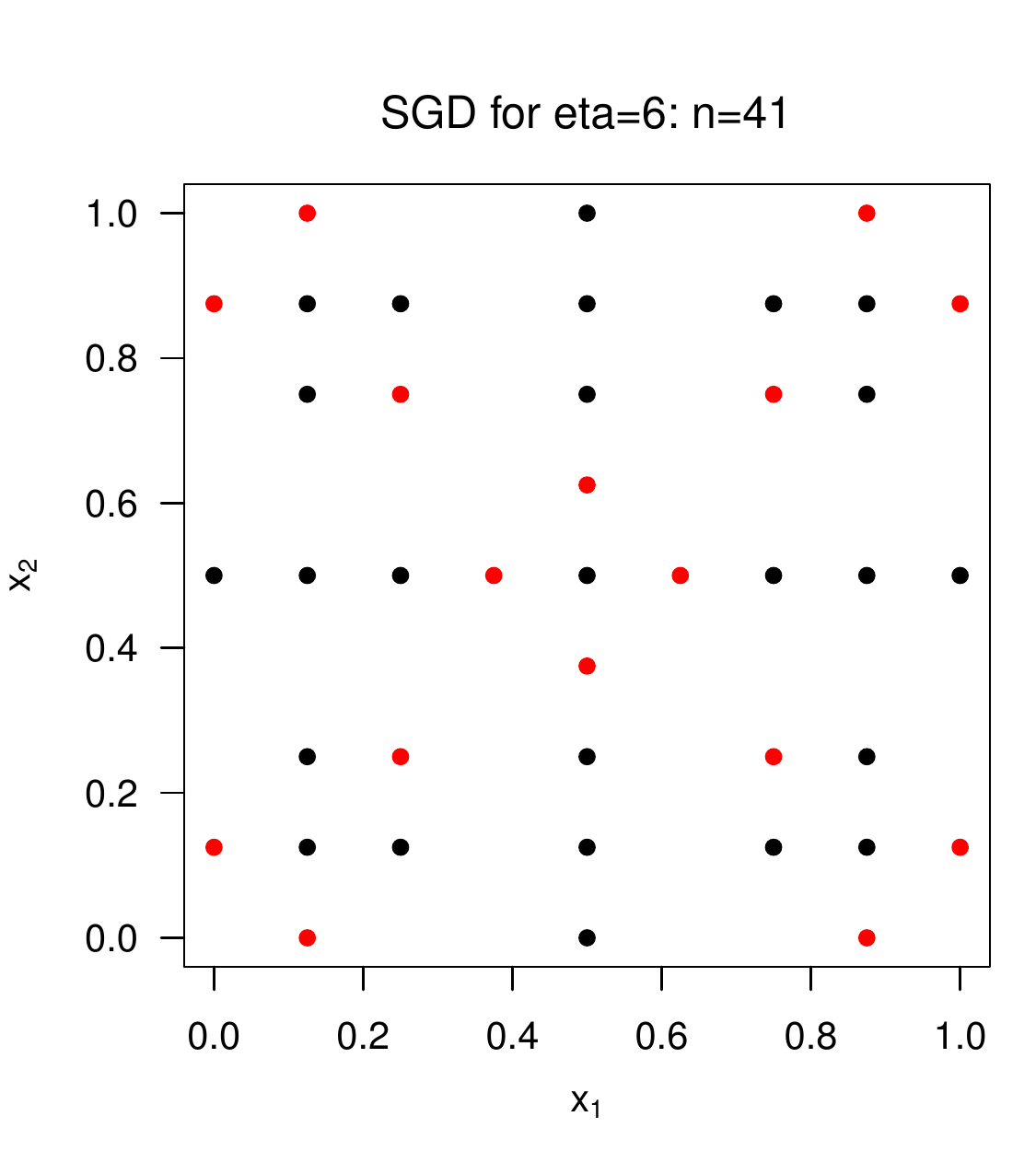}}}\hspace{5pt}
\subfloat{%
\resizebox*{4.55cm}{!}{\includegraphics{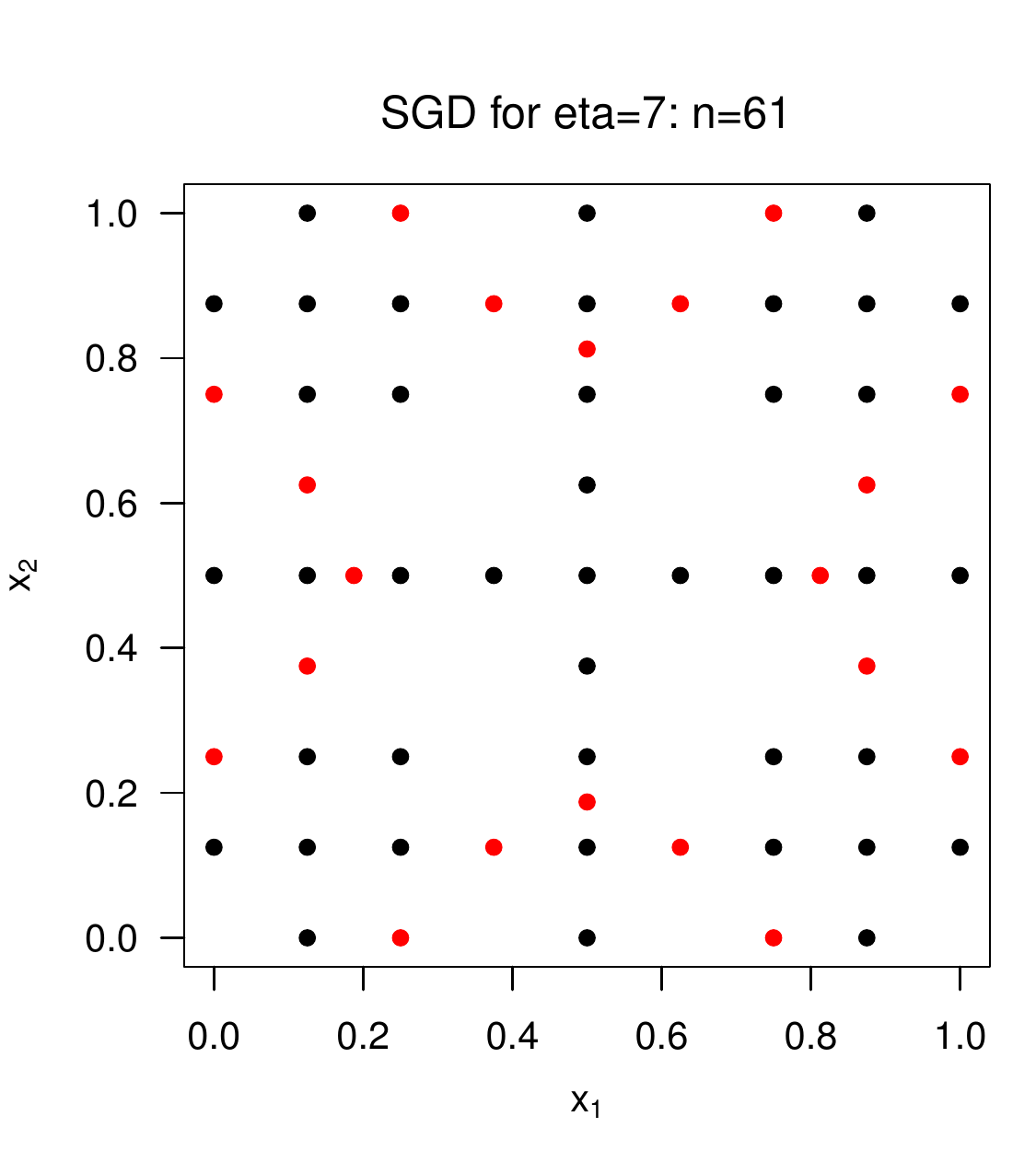}}}\vspace{-2ex}
\caption{Increasing sizes of SGDs in two dimensions. The new points added as $\eta$ increases are denoted by solid red circles.}
\label{fig:sgds}
\end{figure}

SGDs extend the idea of a grid design, while being less sensitive to the curse of dimensionality. The design size is a function of a design parameter $\eta \geq d$, and the SGD is defined as
\begin{equation}
\mathcal{X}(\eta) = \bigcup_{\vec{k} \in \mathbb{G}(\eta)} \mathcal{X}_{1, k_1} \times \mathcal{X}_{2, k_2} \times \hdots \times \mathcal{X}_{d, k_d},
\nonumber
\end{equation}
where each $\mathcal{X}_{j, k}$ is the $k^{\text{th}}$ component design in the $j^{\text{th}}$ dimension, and $\vec{k} = [k_1, \hdots, k_d]^\top$ denotes a set of indices for the component designs. The union is taken over the set $\mathbb{G}(\eta) = \{ \vec{k} \in \mathbb{N}^d : \sum_{j=1}^d k_j = \eta \}$. An additional restriction placed on the SGDs for the purposes of the application is that the component designs must be nested, i.e.~$\mathcal{X}_{j, k} \subseteq \mathcal{X}_{j, k+1}$ for all $j$ and $k$. The first six designs for $d = 2$, constructed using the component designs recommended by \cite{Plumlee2014}, are shown in Figure~\ref{fig:sgds}, with the new points in each design shown in red.

The SGD structure allows for the fast computation of any matrix product $\bm{Q} = \bm{R}^{-1} \bm{M}$, for any $n \times m$ matrix $\bm{M}$, by treating it as a sum of tensor products of linear operators. The code for obtaining the designs and conducting the GP analysis can be found in the MATLAB package \url{Sparse}~\url{Grid}~\url{Designs} \citep{SparseGridDesigns}.




\subsection{Local GP Approximation}
\label{subsec:localgpapproximation}

Local approximate GP (laGP) regression \citep{GramacyApley2015}  is a localization-based method which makes the GP analysis more computationally efficient for an arbitrary design. Prediction at any given location is conducted using a subset of the training data.

Specifically, for prediction at any point $\bm{x}^*$ in the space, the method uses a small subset of the full training design $\mathcal{X} = \{ \bm{x}^{(1)}, \hdots, \bm{x}^{(n)} \}$. The sub-design is defined as some subset $\mathcal{X}_m(\bm{x}^*) \subseteq \mathcal{X}$ containing $m \leq n$ points.

Although a nearest-neighbour strategy would be an intuitive way of choosing the sub-design $\mathcal{X}_m(\bm{x}^*)$, it has been shown that this would be sub-optimal \citep{Steinetal2004}. Instead, \cite{GramacyApley2015} search for $\mathcal{X}_m(\bm{x}^*)$ and the corresponding responses $\bm{y}_m(\bm{x}^*)$, together forming data $\mathcal{D}_m(\bm{x}^*)$, by making a sequence of greedy decisions $\mathcal{D}_{k+1}(\bm{x}^*) = \mathcal{D}_k(\bm{x}^*) \cup (\bm{x}^{(k+1)}, y(\bm{x}^{(k+1)}))$, for iterations $k = n_0, n_0+1, \hdots, m$.

The initial size-$n_0$ sub-design $\mathcal{D}_{n_0}(\bm{x}^*)$ is obtained using the $n_0$ nearest neighbours, for some small number $n_0$. Then, each subsequent $\bm{x}^{(k+1)}$ is chosen to minimize the empirical Bayes mean-squared prediction error (MSPE) for predicting the response at $\bm{x}^*$, given the previous data $\mathcal{D}_k(\bm{x}^*)$.

The result is a computing time of $\mathcal{O}(m^3)$, as opposed to the usual $\mathcal{O}(n^3)$. Furthermore, due to its flexible and localized nature, the resulting model also allows for a non-stationary covariance structure, without any additional computational burden.

Implementation of this method is available in the {\sf R} package \url{laGP} \citep{laGP}. The package allows implementations of either an isotropic correlation structure (i.e.~equal correlation parameters in each dimension), or an anisotropic correlation structure, with the requirement that the correlation must be separable.
  



\subsection{Bayesian Treed GPs}
\label{subsec:bayesiantreedgps}

Proposed by \cite{GramacyLee2008}, Bayesian treed GPs are another localization-based method for fitting GP models using an arbitrary design. Although one of the primary motivations is to create a flexible model that effectively captures non-stationarity and heteroscedasticity, treed GPs also introduce ideas that are useful in solving the computing problem arising from large designs.

The main idea behind treed GPs is that the $d$-dimensional input space is partitioned using a binary tree. Conditional on a given tree, an independent GP model is fitted in each terminal node, i.e.~using only the data falling in the node. Multiple trees are generated by reversible-jump MCMC (RJ-MCMC), in which five possible steps can change a tree: grow, prune, change, swap and rotate \citep{GramacyLee2008}. The method then averages predictions across the generated trees, hence smoothing the different predictive models across the node boundaries of a single tree. It must be noted that this averaging step slows the method down computationally, making it sub-optimal for our purposes.

Implementation of this method is available in the {\sf R} package \url{tgp} \citep{tgp}.




\subsection{Sparse Correlation Matrices}
\label{subsec:sparsecorrelationmatrices}

The final method that we consider here is the one introduced by \cite{Kaufmanetal2011}. The authors propose a model that combines low-order regression terms with compactly supported correlation functions.

Specifically, the usual correlation functions are replaced with their compact counterparts, which truncate to zero for points that are past a certain distance threshold away from each other. In other words, the model assumes that responses at points that are ``far enough" away from each other are essentially uncorrelated. As a result, the correlation matrix $\bm{R}$ becomes sparse, and sparse matrix techniques can be employed to complete the required matrix calculations.

In order to compensate for the information lost in truncating the correlation function, the usual linear (or constant) mean term in \eqref{eq:mu} is modelled more generally, using a linear combination of basis functions which are taken to be tensor products of Legendre polynomials. 

This method is implemented in the {\sf R} package \url{SparseEm}, which is not currently on CRAN but can be found online \citep{SparseEm}.





\section{Adaptive Partitioning Emulator}
\label{sec:adaptivepartitioningemulator}

In this section, we introduce a new approach to fitting a GP to a complex computer model. A function cannot be complex everywhere, otherwise the prediction task is nearly impossible. If there are isolated regions with higher complexity, then a data-adaptive design algorithm may be able to find these regions, and thus avoid having to sample densely across the entire space.  Sequentially acquiring new data based on what has been learned so far is also known as active learning.

Thus, we take a data-adaptive approach to the development of a design, where at each iteration a region in the domain $[0,1]^d$ is split along a chosen dimension, forming two smaller regions, and additional design points are added within each new region. The splits are chosen to reflect the areas of greatest variability and predictive uncertainty within the space.

As the input space is partitioned, a GP regression is fit to each of the new subregions independently. The result is a flexible model that places more focus on the areas of the domain that are more difficult to predict, because those areas have been subject to more attention during the sequential design. An added advantage is the non-stationarity of the resulting model.










\subsection{Methodology}
\label{subsec:methodology}

The APE method is outlined in Algorithm~\ref{algo:ape}. The values of two tuning parameters must be specified beforehand: the size of the initial design $n_0$, and the desired final design size $N$ for stopping.
(Stopping could also be based on a target accuracy measure.)

\begin{algorithm}[t!]
\caption{APE algorithm for GP regression.}
\begin{algorithmic}
\STATE{Initialize the design to be a random $n_0 \times d$ LHD: 
$\mathcal{X} = \{ \bm{x}^{(1)}, \hdots, \bm{x}^{(n_0)} \}$.}
\STATE{Evaluate the responses at this design: $\bm{y} = \{ f(\bm{x}^{(1)}), \hdots, f(\bm{x}^{(n_0)}) \}$.}
\STATE{Define region $\mathcal{R}_1$ as the entire domain $[0,1]^d$.}
\STATE{Find the CV estimate of GP prediction error over $\mathcal{R}_1$: $e_1$.}
\STATE{Set overall design size $n = n_0$ and number of regions $K = 1$.}
\WHILE{($n < N$)}
\STATE{Find region $\mathcal{R}_{k^*}$ with the largest prediction error: $k^* = \text{argmax}_{k = 1, \hdots, K} (e_k)$. \\[-1ex]}
\STATE{Find the number of points that currently lie in $\mathcal{R}_{k^*}$: $n^*$.}
\STATE{Generate a new random $(2 n_0 - n^*) \times d$ LHD within $\mathcal{R}_{k^*}$: $\mathcal{X}^{(\text{new})}$.}
\STATE{Evaluate the responses at this new LHD: $\bm{y}^{(\text{new})}$.}
\STATE{Add the new points to the overall design: $\mathcal{X} \leftarrow \{ \mathcal{X}, \bm{x}^{(\text{new})} \}$.}
\STATE{Add the new responses to the overall response set: $\bm{y} \leftarrow \{ \bm{y}, \bm{y}^{(\text{new})} \}$.}
\STATE{In $\mathcal{R}_{k^*}$:}
\STATE{\hspace{2ex} Choose dimension for splitting: $j^*$ (see Algorithm~\ref{algo:choosedim}).}
\STATE{\hspace{2ex} Denote the mid-point of $x_{j^*}$ in this region by $\tilde{x}_{j^*, k^*}$.}
\STATE{\hspace{2ex} Split at $\tilde{x}_{j^*, k^*}$, forming two new regions, $\mathcal{R}_{\text{new}, 1}$ and $\mathcal{R}_{\text{new}, 2}$.}
\FOR{$\ell$ in $1:2$}
\STATE{Train a GP using the data in $\mathcal{R}_{\text{new}, \ell}$.}
\STATE{Find the CV estimate of GP prediction error in $\mathcal{R}_{\text{new}, \ell}$. Denote by $e_{\text{new}, \ell}$.}
\ENDFOR
\STATE{Then, $\mathcal{R}_{\text{new}, 1}$ replaces $\mathcal{R}_{k^*}$, and $\mathcal{R}_{\text{new}, 2}$ becomes $\mathcal{R}_{K+1}$.}
\STATE{Similarly, $e_{\text{new}, 1}$ replaces  $e_{k^*}$, and $e_{\text{new}, 2}$ becomes $e_{K+1}$.}
\STATE{Update the total number of regions:  $K \leftarrow K + 1$.}
\STATE{Update the overall design size:  $n \leftarrow n + (2n_0 - n^*)$.}
\ENDWHILE
\end{algorithmic}
\label{algo:ape}
\end{algorithm}

\begin{algorithm}
\caption{Choosing the splitting dimension, within a region $\mathcal{R}_{k^*}$.}
\begin{algorithmic}
\FOR{dimension $j$ in $1:d$}
\STATE{Propose a split in dimension $j$.}
\STATE{Denote the mid-point of $x_j$ in the region $\mathcal{R}_{k^*}$ by $\tilde{x}_{j, k^*}$.}
\STATE{Split at $\tilde{x}_{j, k^*}$, forming two new hypothetical sub-regions.}
\FOR{hypothetical sub-region $\ell$ in $1:2$}
\STATE{Find the mean of the responses within sub-region $\ell$: $m_j^{(\ell)}$.}
\STATE{Find the variance of the responses within sub-region $\ell$: $v_j^{(\ell)}$.}
\ENDFOR
\STATE{Calculate the variance of the means: $V_{j, \text{between}} = \sum_{\ell=1}^2 \left( m_j^{(\ell)} - \bar{m}_j \right)^2$, \\ \hspace{3ex} where $\bar{m}_j = \frac{1}{2} \sum_{\ell=1}^2 m_j^{(\ell)}$.}
\STATE{Calculate the mean of the variances: $V_{j, \text{within}} = \frac{1}{2} \sum_{\ell=1}^2 v_j^{(\ell)}$.}
\ENDFOR
\STATE{Choose the dimension for splitting, $j^* = \text{argmin}_{j = 1, \hdots, d} (V_{j, \text{within}} / V_{j, \text{between}})$.}
\end{algorithmic}
\label{algo:choosedim}
\end{algorithm}

The algorithm begins with a single region equal to the entire input domain $[0,1]^d$.
The design for this initial region is a random $n_0 \times d$ 
Latin hypercube design \citep[LHD,][]{McKBecCon1979} $\mathcal{X} = \{ \bm{x}^{(1)}, \hdots, \bm{x}^{(n_0)} \}$, where each $\bm{x}^{(i)}$ is a $d$-dimensional point in $[0,1]^d$. The response vector $\bm{y}$ for this design is obtained by evaluating the computer model $f(\cdot)$ at each of the design points in $\mathcal{X}$.

An estimate of the prediction error $e_1$ over the initial region is then obtained using leave-one-out cross validation (CV), whereby one observation is omitted from the dataset at a time, and the remaining observations are used to fit a GP model to predict that observation. The prediction error is evaluated as the mean squared error (MSE) of the leave-one-out errors. Alternatively, the maximum absolute error can be used instead.

The algorithm then begins its first iteration and continues until a specified stopping rule is met. The stopping rule chosen here is simply that the overall design size has reached or surpassed a desired value $N$. Alternatively, the algorithm could be set to terminate after a fixed number of iterations, or once a given error tolerance is reached.

At each iteration, the algorithm proceeds as follows. First, the region with the largest prediction error is identified. At the first iteration, before any splitting has occurred, this is simply the entire domain $[0,1]^d$. The chosen region is then denoted by the index $k^*$. This is the region that will be split at this iteration of the algorithm.

Before the splitting occurs, the design size in region $k^*$ is increased to be equal to $2 n_0$, by adding a new LHD of size $(2 n_0 - n^*)$, where $n^*$ is the current number of points in the region.

The next step in the algorithm is the choice of the dimension along which the splitting will occur. The dimension is chosen such that a potential split along this dimension would minimize the ratio of within-region variance to between-region variance of the two resulting sub-regions. The justification is that it is desirable for the new sub-regions to be both homogeneous and different from each other, in order for the split to be meaningful. Moreover, the computation is very quick in contrast to, say, fitting GPs for every candidate splitting dimension. Full details of this calculation are outlined in Algorithm~\ref{algo:choosedim}.

Once the region $k^*$ and dimension $j^*$ have been chosen, the region is then split along this dimension, at the midpoint $\tilde{x}_{j^*,k^*}$, forming two new regions.

The next step involves evaluating the performance of the GP model in each of the two new regions, using a leave-one-out CV in each region independently. This is done for the purpose of choosing which of the $K$ regions will be split at the next iteration. Note that the CV estimates of prediction error for all previously existing regions would have been calculated already, at previous iterations, and thus do not need to be calculated again.

The algorithm terminates once the overall design size $N$ is reached or surpassed. The result is a set of regions partitioning the domain, chosen sequentially, where each region is expected to contain an approximately equal number of observations. A GP regression is fit to each of the regions, resulting in flexible non-stationary (across regions) models from data that place more focus on areas of the domain with greater variability.





\section{Results}
\label{sec:results}

The performance of the above methods is compared using two test functions of different complexity and dimensionality. Code and detailed descriptions for the functions can be found in the online Virtual Library of Simulation Experiments \citep{SurjanovicBingham2013}.

We are interested in comparing the performance of the following methods for each test function: the standard GP fitting method \citep[using the \url{mlegp} package in {\sf R}; ][]{mlegp}, Sparse Grid Designs (SGD), local GP approximation with a separable covariance function (laGPsep), Bayesian treed GPs (tgp), and sparse correlation matrices (SparseEm).

For each test function, we construct SGDs of varying sizes, to serve as training sets for the SGD method. For each of these design sizes $n$, we also construct a random $n \times d$ Latin Hypercube Design (LHD) to serve as a training set for the other existing methods (mlegp, laGPsep, tgp and SparseEm). Note that the same LHDs are used across each of these methods, in order to reduce variability due to sampling error. Designs for the APE method are obtained sequentially, as described in Algorithm~\ref{algo:ape}.

Each of the methods is then used to construct a fitted model using the appropriate training set, and predictions are evaluated on a test set $\{ y^*_1, \hdots, y^*_{n_\text{test}} \}$, obtained at $n_{\text{test}} = 10\,000$ random points uniformly sampled across the space. Note that the same test set is used across each of the methods and design sizes.

Predictive performance of each method is evaluated using two different measures. The root mean squared prediction error (RMSPE), for a given method and a given design size $n$, is evaluated as
\begin{equation}
\text{RMSPE}= \sqrt{\frac{1}{n_{\text{test}}} \sum_{i=1}^{n_{\text{test}}} (y^*_i - \hat{y}^*_i)^2},
\nonumber
\end{equation}
where, in this case, $\{ \hat{y}^*_1, \hdots, \hat{y}^*_{n_{\text{test}}} \}$ are the predictions obtained from this method and this design size specifically. For scenarios in which a ``worst-case" error measure is of interest, the maximum absolute prediction error (MAPE) is given by
\begin{equation}
\text{MAPE} = \max_{i = 1, \hdots, n_{\text{test}}} \big| y^*_i - \hat{y}^*_i \big|.
\nonumber
\end{equation}

In order to make the above error measures easier to interpret, they are scaled by a measure of the variability of the test function. Specifically, RMSPE is divided by the standard deviation of the test set, while MAPE is divided by its maximum absolute deviation from the mean.

Finally, the elapsed computing time, in minutes, is also given for each method and each design size.




\subsection{Corner Peak Function}
\label{subsec:cornerpeakfunction}

\begin{figure}
\begin{center}
\includegraphics[width=0.6\textwidth]{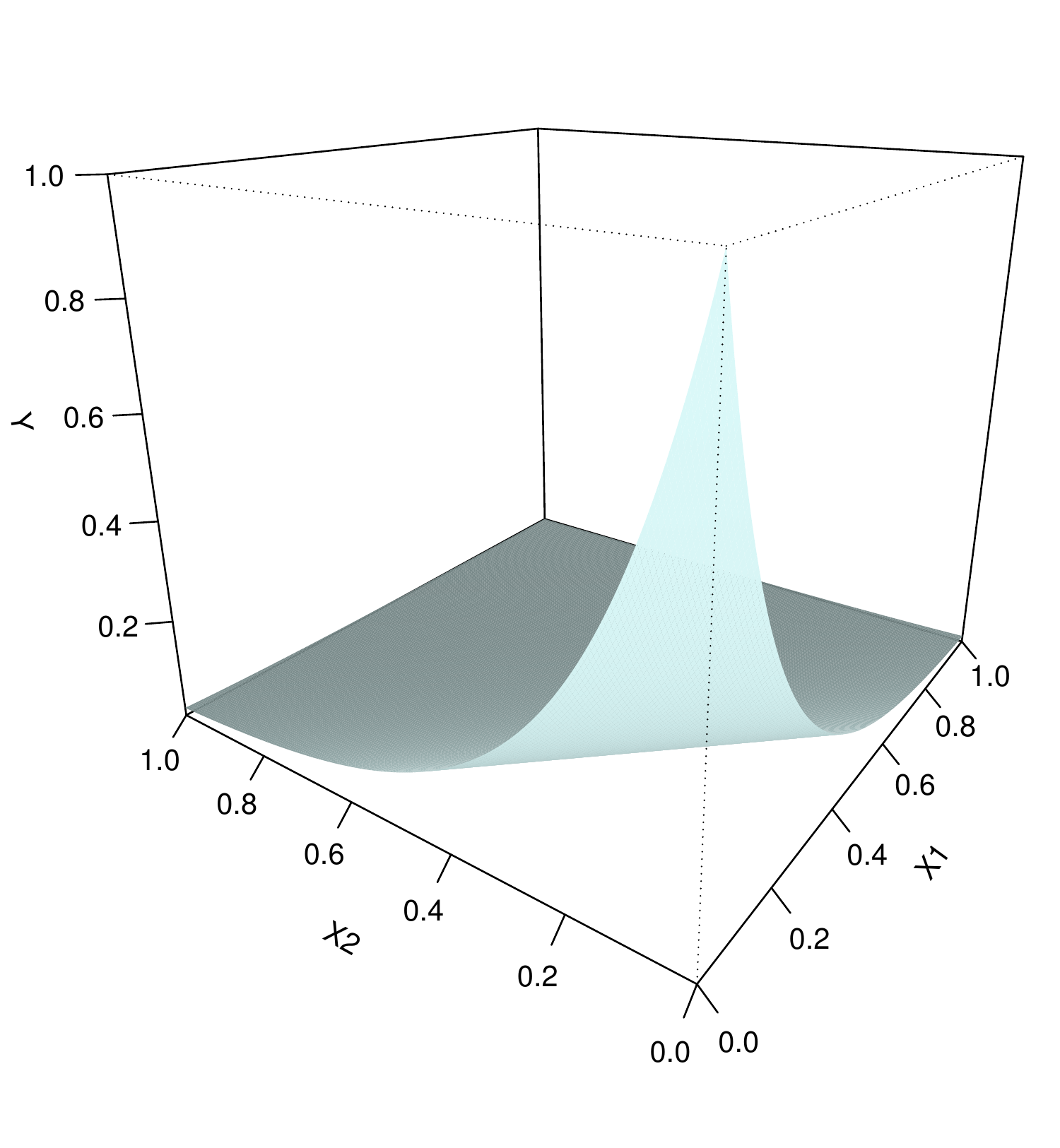}
\vspace{-5ex}
\caption{First two dimensions of the corner peak function.}
\label{fig:cornerpeak}
\end{center}
\end{figure}

The first function we use to demonstrate the performance of the methods is the corner peak function:
\begin{equation}
f({\bm x}) = \left( 1 + \sum_{j=1}^d a_j x_j \right)^{-(d+1)},
\label{eq:cornerpeak}
\end{equation}
which is defined on $x_j \in [0, 1]$ and $a_j > 0$, for $j = 1, \hdots, d$.

This function is characterized by a fairly flat surface that rises in a sharp peak near the origin. Specifically, a larger value of the $a_j$ parameter in a given dimension results in a steeper rise in the peak in that dimension, thus resulting in more difficult emulation and prediction.

In order to achieve a moderate level of difficulty in $d = 10$ dimensions, \cite{Barthelmannetal2000} suggest uniformly sampling the $a_j$ parameters from 0 to 1, and then rescaling them such that $\sum_{j=1}^{10} a_j = 1.85$. This approach is also followed by \cite{Chen2018}. For the purposes of future comparison to results found therein, we use the same random parameter values: $\{ 0.4761, 0.4500, 0.3297, 0.2553, 0.0963, 0.0764, 0.0714, 0.0648, 0.0286, 0.0014 \}$, where the values have been sorted largest-to-smallest without loss of generality. The first two dimensions of the resulting function are shown in Figure~\ref{fig:cornerpeak}, with all other $x_j$-values set to zero.

The first three SGDs in 10 dimensions are of sizes 21, 221 and 1561. We obtain these SGDs, for use as a training set for the SGD method. For the LHD-based methods, we construct random LHDs with the same sizes, as well as an additional LHD of size 3200 for further comparison. The corner peak function is then evaluated at all of the resulting training sets. Each of the methods in Section~\ref{sec:gaussianprocessregressionwithlargedatasets} is used to construct a fitted model using the appropriate training set, and to make predictions at the 10\,000 points in the test set. 

\begin{figure}
\begin{center}
\includegraphics[width=0.6\textwidth]{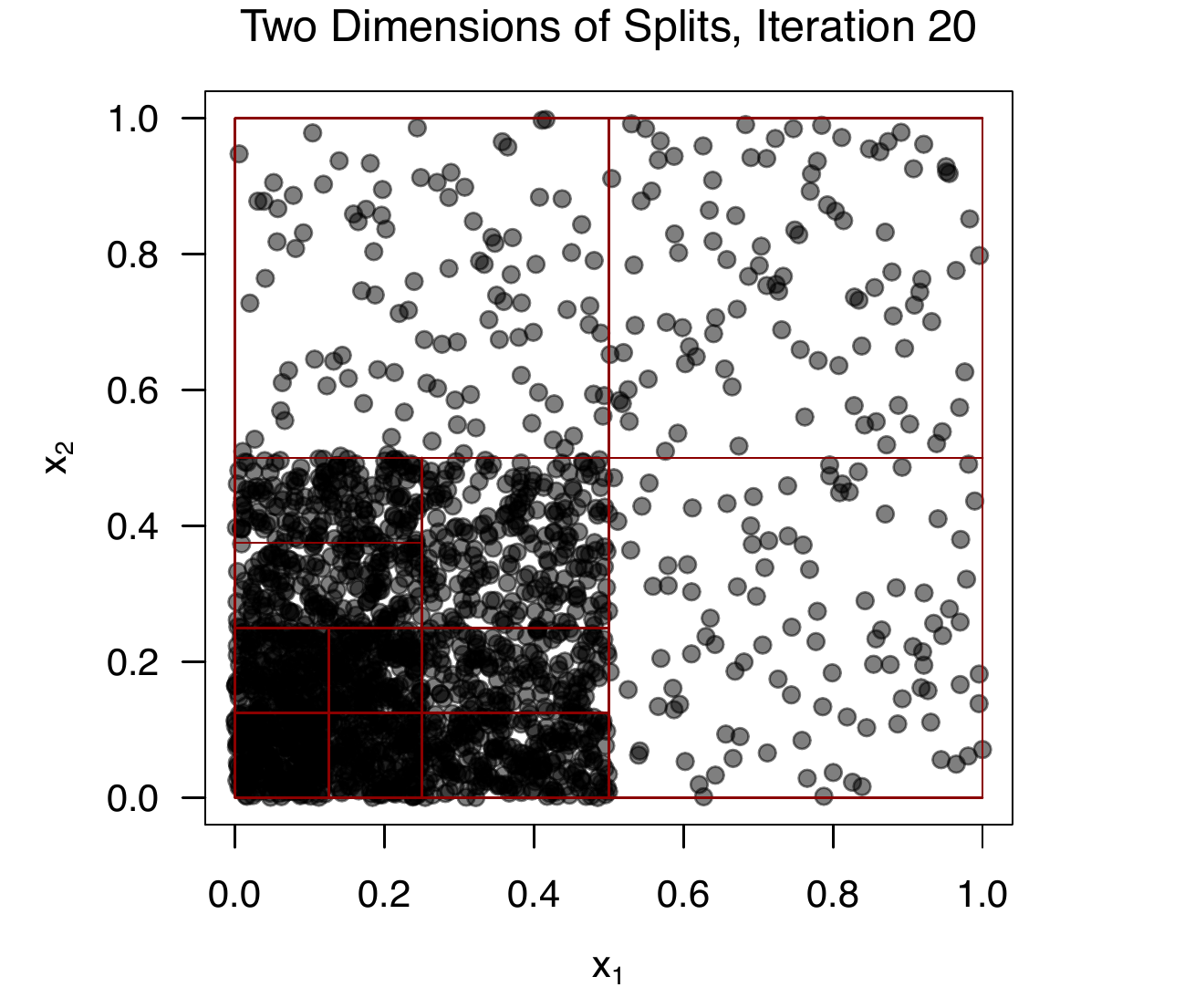}
\vspace{-3ex}
\caption{Splits in the first two dimensions made by the APE algorithm at iteration 20 for the corner peak function, with $n_0 = 100$.}
\label{fig:cornerpeakregions}
\end{center}
\end{figure}

For comparison, we also run the APE algorithm with two different values of $n_0$, the number of initial observations chosen on a random LHD: $100$ and $200$. In each case, the algorithm is run up to a maximum design size of $4000$. The resulting regions formed after the first 20 iterations (for $n_0 = 100$) are shown in Figure~\ref{fig:cornerpeakregions}. As this function is very steep at the origin, the sequential design algorithm is able to detect the sharp peak and place more points near that corner.

\begin{figure}
\centering
\subfloat{%
\resizebox*{15cm}{!}{\includegraphics{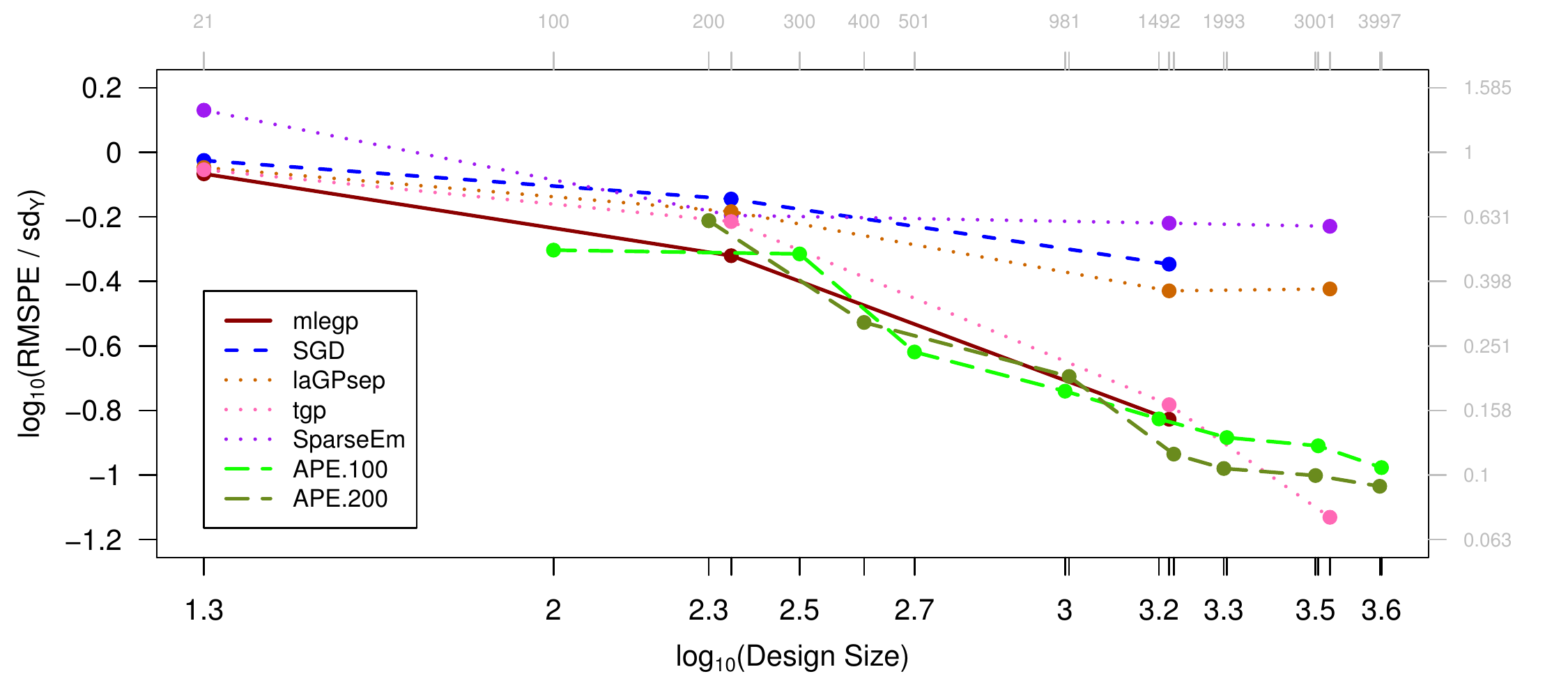}}}\hspace{5pt}
\subfloat{%
\resizebox*{15cm}{!}{\includegraphics{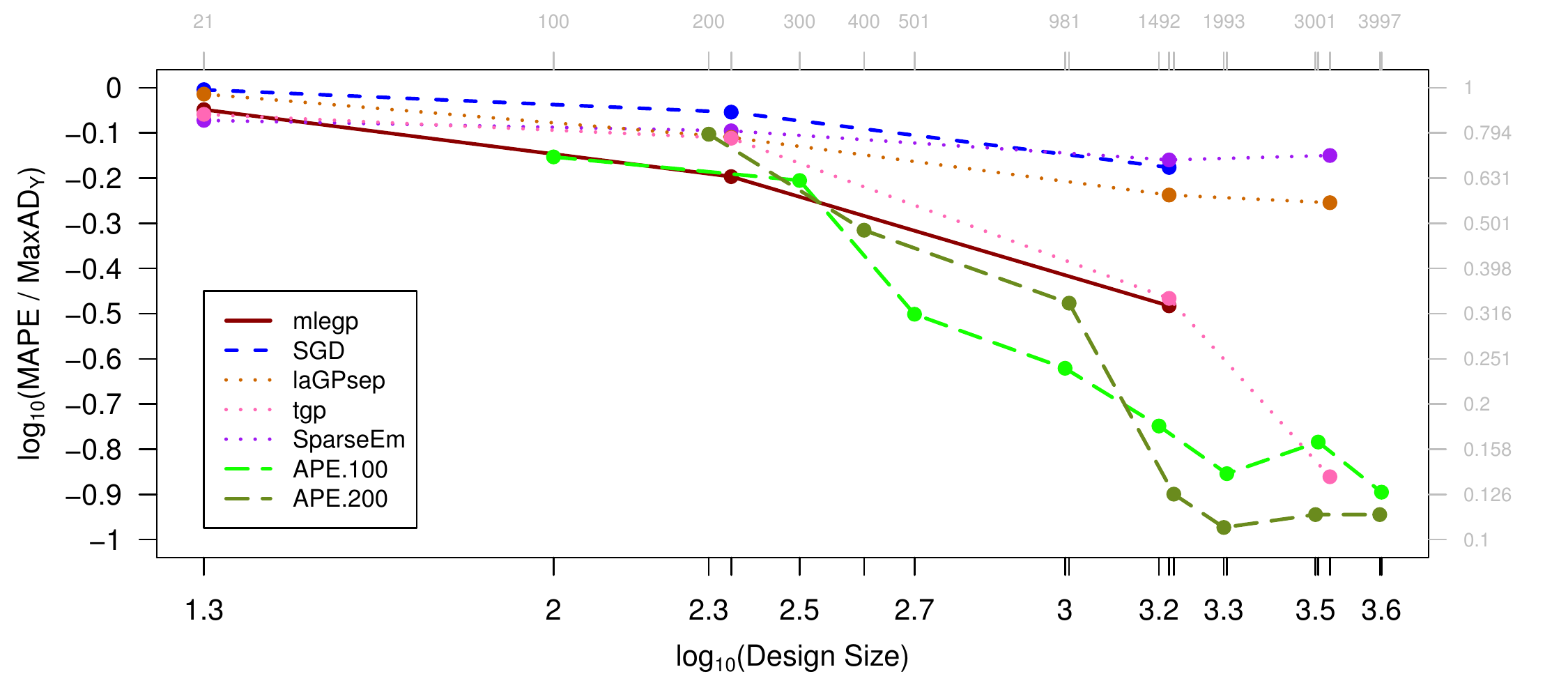}}}\hspace{5pt}
\subfloat{%
\resizebox*{15cm}{!}{\includegraphics{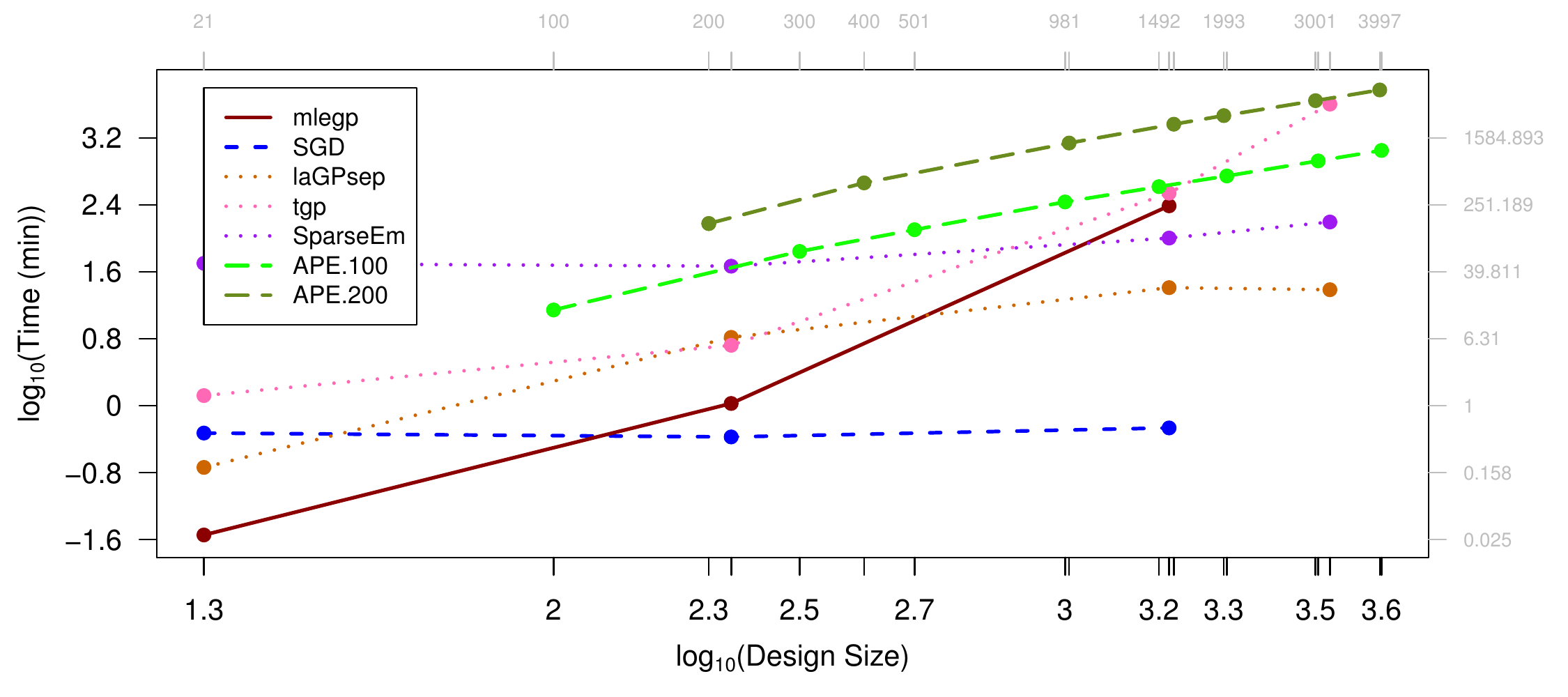}}}\vspace{-2ex}
\caption{Scaled RMSPE (top), scaled MAPE (middle) and time (bottom) versus design size (log-log scales)
for the corner peak function. APE.100 is the APE algorithm with $n_0 = 100$.}
\label{fig:cornerpeakplots}
\end{figure}

The results for all methods are shown in Figure~\ref{fig:cornerpeakplots}. The plots show scaled RMSPE (denoted by $\text{RMSPE} / \text{sd}_Y$), scaled MAPE (denoted by $\text{MAPE} / \text{MaxAD}_Y$), and computing time, against the design size, on a log-log scale. For ease of use, unlogged values are given in grey on the opposite axes. Tables~\ref{tab:cornerpeakresults}~and~\ref{tab:cornerpeakresults2} also present all of the unlogged values.

\renewcommand{\tabcolsep}{0.9em}
\renewcommand{\arraystretch}{1}
\begin{center}
\begin{table}[t!]
\center
\caption{Results for the corner peak function: scaled RMSPE (black), scaled MAPE (blue), and time in minutes (red).}
  \begin{tabular}{|=c|+r|+r|+r|+r|}
    \hline
    \multicolumn{1}{|>{\columncolor{white}}c|}{} & \multicolumn{4}{c|}{Design Size}\\
    \arrayrulecolor{kugray5}
    \arrayrulecolor{black}
    \cline{2-5}
    \multicolumn{1}{|>{\columncolor{white}}c|}{} & 21        & 221      & 1561     & 3200\\
    \hline
    \multirow{3}{*}{mlegp}                                   & 0.858   & 0.478   & 0.149    & \textcolor{black}{--} \\
                                    \rowstyle{\color{blue2}} & 0.895   & 0.636   & 0.329    & \textcolor{black}{--} \\
                                      \rowstyle{\color{red2}} & 0.028   & 1.063   & 244.721& \textcolor{black}{--} \\
    \hline
    \multirow{3}{*}{SGD}                                     & 0.943   & 0.716   & 0.451     & \textcolor{black}{--} \\
                                    \rowstyle{\color{blue2}} & 0.990   & 0.883   & 0.666     & \textcolor{black}{--} \\
                                      \rowstyle{\color{red2}} & 0.471   & 0.423   & 0.542    & \textcolor{black}{--} \\
    \hline
    \multirow{3}{*}{laGPsep}                               & 0.897   & 0.655   & 0.372     & 0.377 \\
                                    \rowstyle{\color{blue2}} & 0.968   & 0.776   & 0.579     & 0.557 \\
                                      \rowstyle{\color{red2}} & 0.183   & 6.559   & 25.739  & 24.325 \\
    \hline
    \multirow{3}{*}{tgp}                                        & 0.883   & 0.610   & 0.165     &0.074 \\
                                    \rowstyle{\color{blue2}} & 0.873   & 0.773   & 0.342     & 0.138 \\
                                      \rowstyle{\color{red2}} & 1.321   & 5.268   & 347.266 & 4021.858 \\
    \hline
    \multirow{3}{*}{SparseEm}                            & 1.350   & 0.638   & 0.603      & 0.590 \\
                                    \rowstyle{\color{blue2}} & 0.847   & 0.803   & 0.692      & 0.708 \\
                                      \rowstyle{\color{red2}} & 50.329 & 46.519 & 100.764  & 156.822\\
    \hline
  \end{tabular}
\label{tab:cornerpeakresults}
\end{table}
\end{center}

\renewcommand{\tabcolsep}{0.9em}
\renewcommand{\arraystretch}{1}
\begin{center}
\begin{table}[t!]
\center
\caption{APE results for the corner peak function: scaled RMSPE (black), scaled MAPE (blue), and time in minutes (red).}
  \begin{tabular}{|=c|+r|+r|+r|+r|+r|+r|+r|+r|}
    \hline
    \multicolumn{1}{|>{\columncolor{white}}c|}{} & \multicolumn{6}{c|}{Design Size}\\
    \arrayrulecolor{kugray5}
    \arrayrulecolor{black}
    \cline{2-7}
    \multicolumn{1}{|>{\columncolor{white}}c|}{} & \textcolor{black}{501} & \textcolor{black}{981} & \textcolor{black}{1492} & \textcolor{black}{2020} & \textcolor{black}{3037} & \textcolor{black}{4029} \\
    \hline
    \multirow{3}{*}{$n_0=100$}             & 0.241  & 0.182   & 0.149   & 0.131   & 0.123 & 0.105 \\
                       \rowstyle{\color{blue2}} & 0.315  & 0.239   & 0.178   & 0.140   & 0.164 & 0.127 \\
                        \rowstyle{\color{red2}} & 126.83 & 272.50 & 414.86 & 557.00 & 842.52 & 1124.63 \\
    \hline
   \multicolumn{7}{|>{\columncolor{kugray5}}c|}{} \\
   \hline
    \multicolumn{1}{|>{\columncolor{white}}c|}{} & \multicolumn{6}{c|}{Design Size}\\
    \cline{2-7}
    \multicolumn{1}{|>{\columncolor{white}}c|}{} & \textcolor{black}{400} & \textcolor{black}{999} & \textcolor{black}{1594} & \textcolor{black}{1993} & \textcolor{black}{3001} & \textcolor{black}{3997} \\
   \hline
    \multirow{3}{*}{$n_0=200$}             & 0.297  & 0.202     & 0.116      & 0.105     & 0.100     & 0.092 \\
                       \rowstyle{\color{blue2}} & 0.484  & 0.334     & 0.126     & 0.106     & 0.114     & 0.114 \\
                        \rowstyle{\color{red2}} & 460.13 & 1378.32 & 2314.23 & 2932.23 & 4426.20 & 5939.89 \\
    \hline

  \end{tabular}
\label{tab:cornerpeakresults2}
\end{table}
\end{center}

\newpage 
The standard GP fitting method (``mlegp") and tgp outperform laGPsep, SGD, and SparseEm here in terms of predictive ability. However, the computing time for both methods drastically increases for larger design sizes. The rate of increase with respect to design size is much higher than for the other methods.
With mlegp, this is due to the computational complexity behind the usual GP fitting calculations; so much so that the computation became intractable for $n = 3200$. With tgp, most of the computing time lies in the step that involves ``averaging" over all possible trees.

In contrast, SGD, laGPsep and SparseEm have somewhat weaker predictive ability, but a much lower rate of increase in computing time with respect to the design size. In particular, SparseEm and SGD appear to have a near constant running time. However, it is debatable whether this advantage is beneficial for moderate design sizes here, due to the loss in predictive ability.

Finally, we see that APE is competitive with the other methods. In terms of predictive ability, it performs similarly to mlegp and tgp, with $n_0 = 200$ slightly outperforming $n_0 = 100$. The overall long computing times could be improved by better implementation of APE; most importantly, it is clear that the rate of increase with respect to the design size is much lower than for mlegp and tgp.




\subsection{Four-Dimensional Franke Function}
\label{subsec:fourdimensionalfrankefunction}

\begin{figure}
\begin{center}
\includegraphics[width=0.6\textwidth]{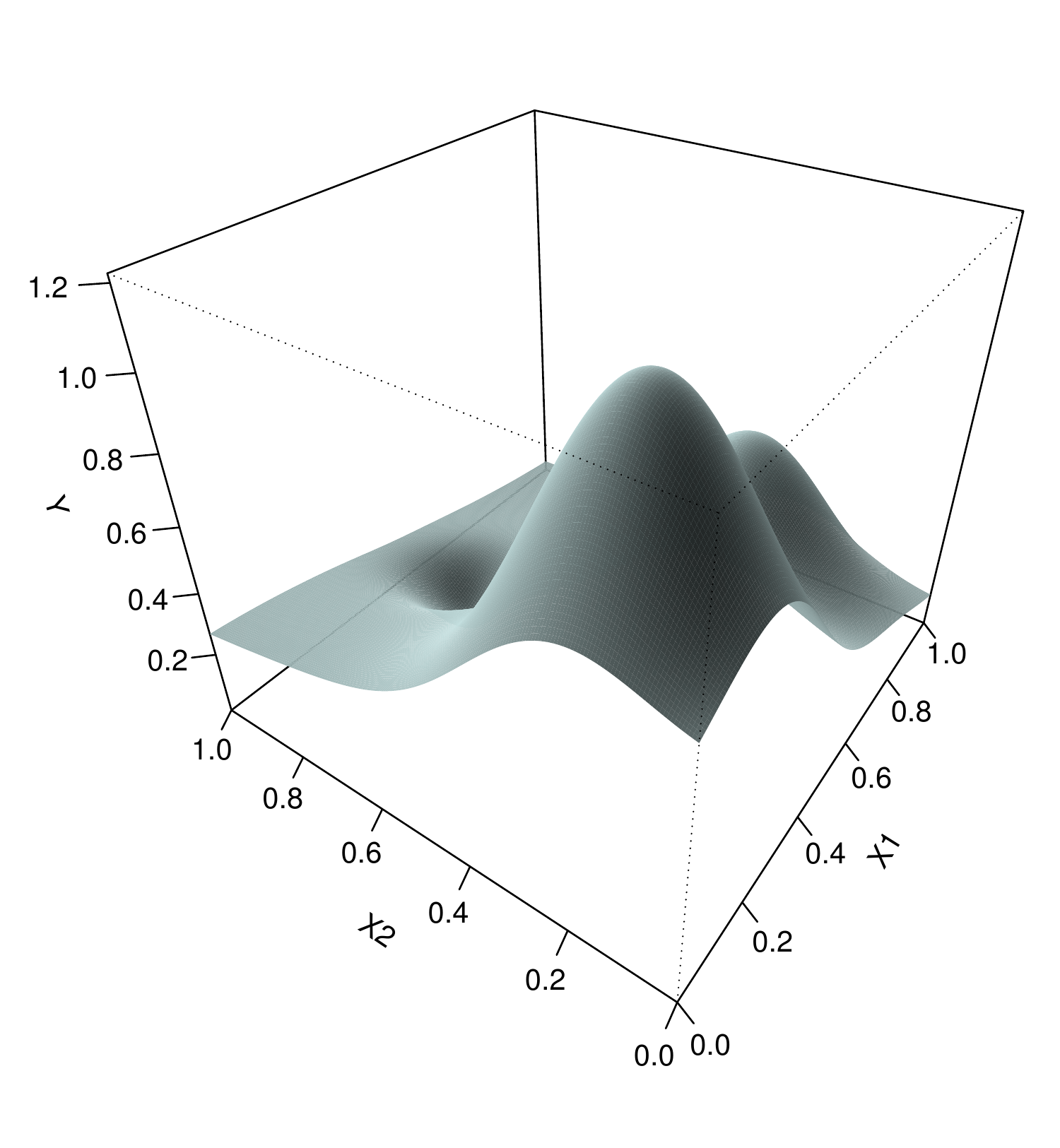}
\vspace{-4ex}
\caption{Bivariate Franke function.}
\label{fig:franke2d}
\end{center}
\end{figure}

The bivariate Franke function \citep{Franke1979}, originally proposed for interpolation problems, is 
\begin{align}
g(x_1, x_2) &= 0.75 \, \text{exp}\left( - \frac{(9 x_1 - 2)^2}{4} - \frac{(9 x_2 - 2)^2}{4} \right) 
\nonumber \\
&+ 0.75 \, \text{exp}\left( - \frac{(9 x_1 + 1)^2}{49} - \frac{(9 x_2 + 1)^2}{10} \right) \nonumber \\
&+ 0.5 \,\text{exp}\left( - \frac{(9 x_1 - 7)^2}{4} - \frac{(9 x_2 - 3)^2}{4} \right) \nonumber \\
&-0.2 \, \text{exp}\left( - (9 x_1 - 4)^2 - (9 x_2 - 7)^2 \right). \label{eq:franke2d}
\end{align}
The inputs are defined on $[0,1]^2$. As shown in Figure~\ref{fig:franke2d}, this function is characterized by two Gaussian peaks of different heights, and a smaller dip. Due to its smoothness and low dimensionality, it is fairly easily modelled using a Gaussian process regression.

In order to make the problem slightly more challenging, we formulate a four-dimensional version of the Franke function as a sum of two marginal bivariate Franke functions:
\begin{equation}
f({\bm x}) = g(x_1, x_2) + g(x_3, x_4),
\label{eq:franke4d}
\end{equation}
where each $g(\cdot, \cdot)$ is defined as in \eqref{eq:franke2d}.

The first eight SGDs in four dimensions are of sizes 9, 41, 129, 321, 681, 1289, 2241 and 3649. We obtain these SGDs, as well as random LHDs of the same sizes, and evaluate the four-dimensional Franke function at each of the points. Each of the methods in Section~\ref{sec:gaussianprocessregressionwithlargedatasets} is used to construct a fitted model using the appropriate training set, and to make predictions at the 10\,000 points in the test set. 

\begin{figure}
\begin{center}
\includegraphics[width=0.6\textwidth]{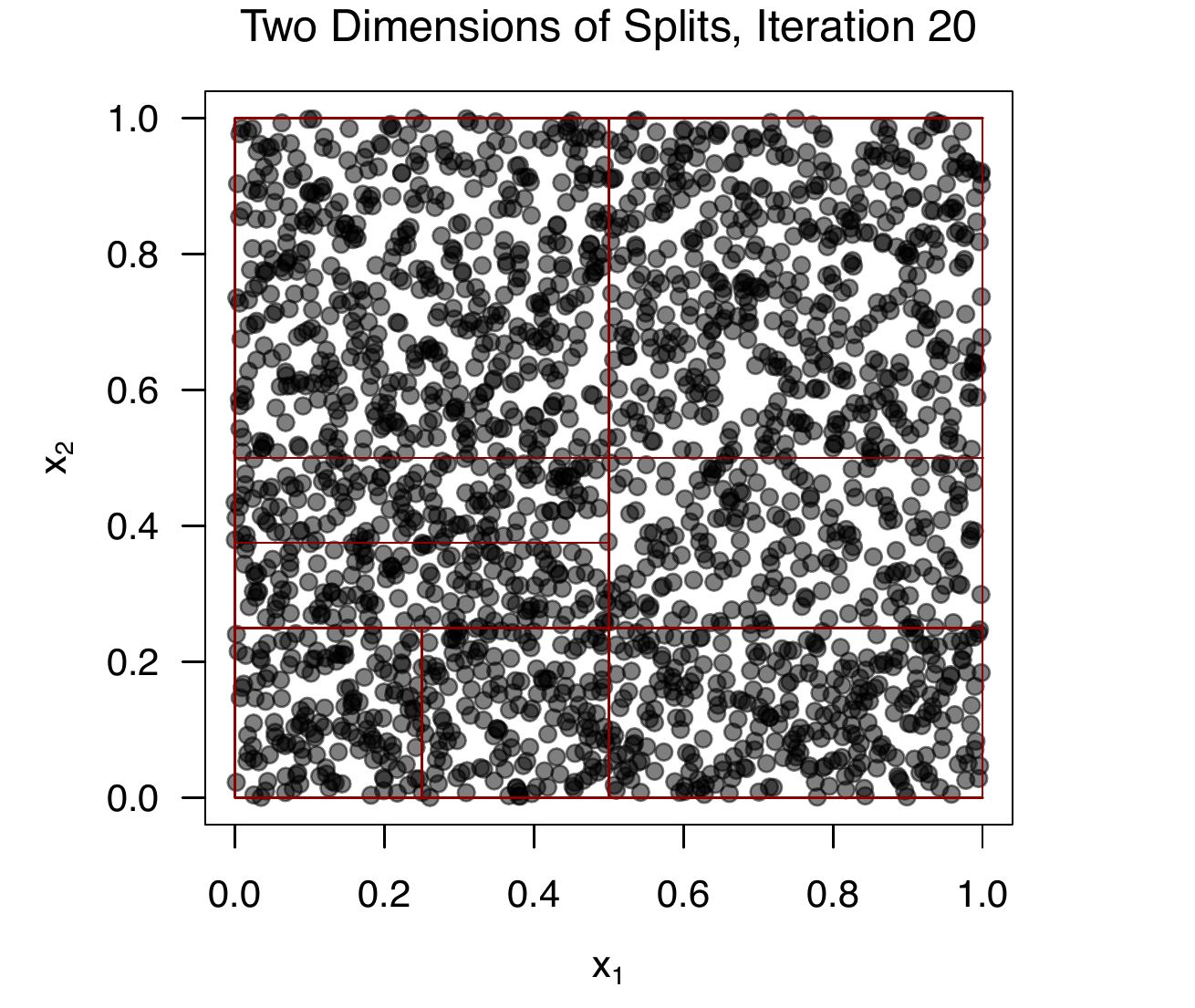}
\vspace{-3ex}
\caption{Splits in the first two dimensions made by the APE algorithm at iteration 20 for the four-dimensional Franke function, with $n_0 = 100$.}
\label{fig:franke4dregions}
\end{center}
\end{figure}

We run the APE algorithm for $n_0 = 100$, up to a maximum design size of $3600$. The resulting regions formed for $x_1$ and $x_2$ after the first 20 iterations are shown in Figure~\ref{fig:franke4dregions} 
(other splits among the first 20 occur in the third and fourth dimensions). 
Since this function's variability is much more constant across the domain, splitting occurs much more evenly across the dimensions, though there is more emphasis towards the origin where the larger Gaussian peak is located.

\begin{figure}[t!]
\centering
\subfloat{%
\resizebox*{15cm}{!}{\includegraphics{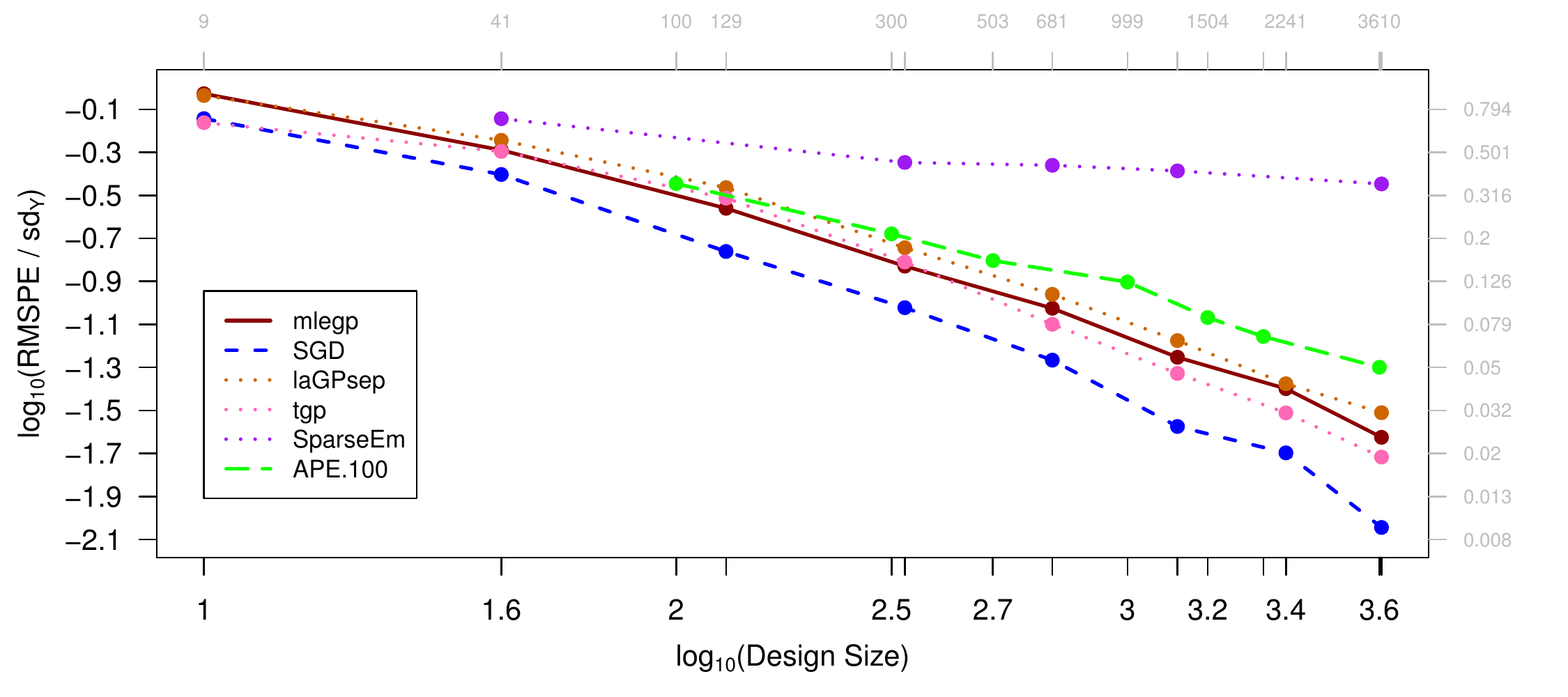}}}\hspace{5pt}
\subfloat{%
\resizebox*{15cm}{!}{\includegraphics{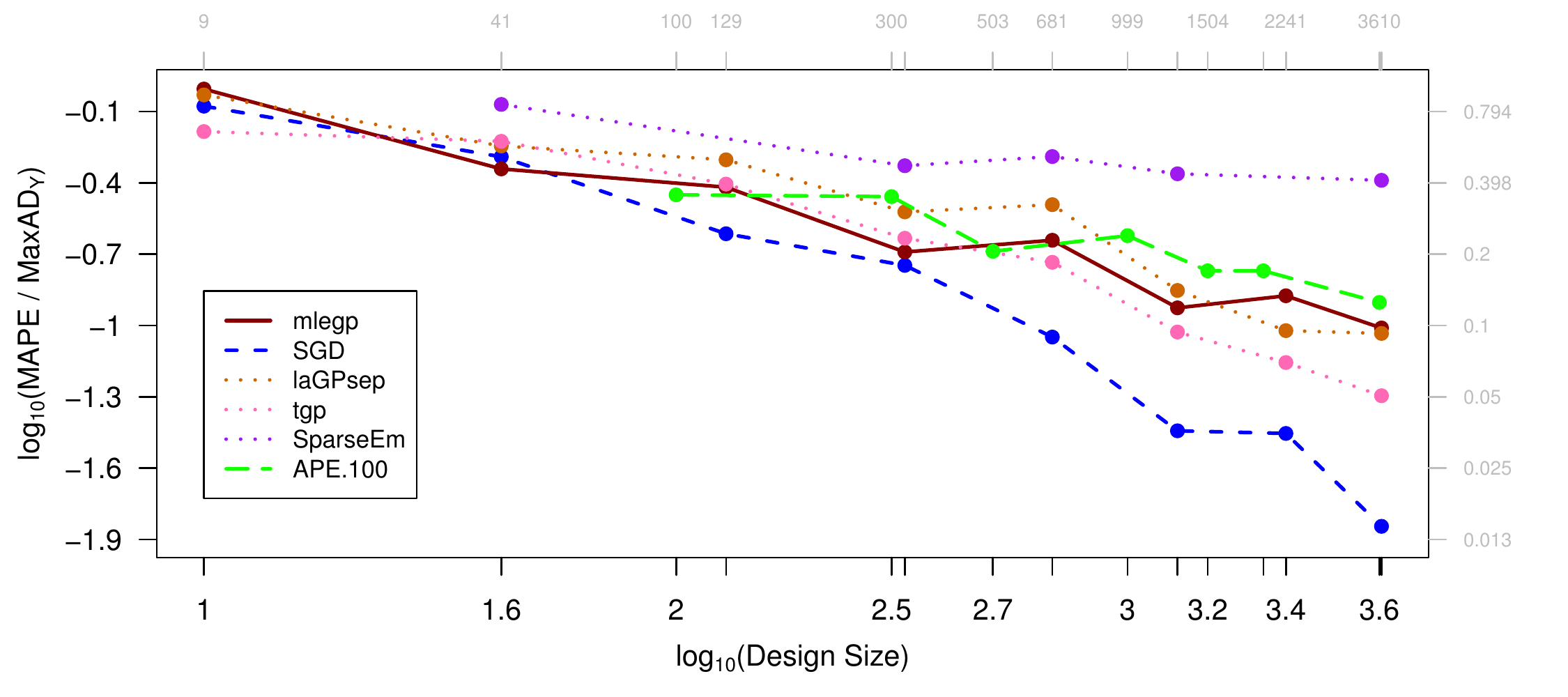}}}\hspace{5pt}
\subfloat{%
\resizebox*{15cm}{!}{\includegraphics{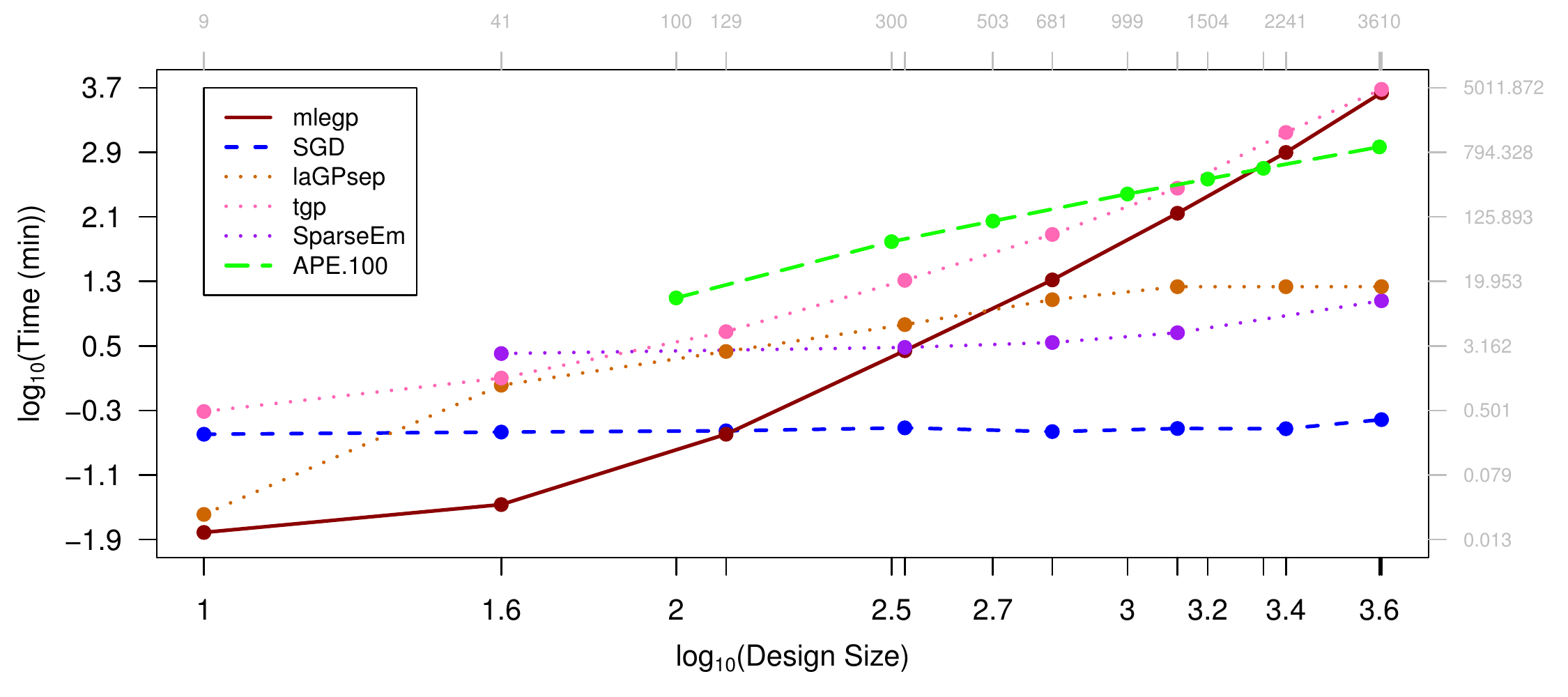}}}\vspace{-2ex}
\caption{Scaled RMSPE (top), scaled MAPE (middle) and time (bottom) versus design size (log-log scales)
for the four-dimensional Franke function. APE.100 is the APE algorithm with $n_0 = 100$.}
\label{fig:franke4dplots}
\end{figure}

The results of the simulation study are shown in Figure~\ref{fig:franke4dplots}. As before, the plots show scaled RMSPE, scaled MAPE, and computing time, against the design size, on a log-log scale. The unlogged values are given on the opposite axes, as well as in Tables~\ref{tab:franke4dresults}~and~\ref{tab:franke4dresults2}.

As before, we see that the computing time for the mlegp and tgp methods increases much more quickly with design size than for the other methods. For the largest design size, the computing time for these two methods was more than 3 days each. However, the results for predictive performance are very different for this function than for the corner peak function. Due to the function's smoothness and fairly low dimensionality, it is easily fitted using almost all of the methods. As a result, the scaled RMSPE-values are very similar. There is slightly more variability in the scaled MAPE-values, but this is likely due to the higher inherent variability of a maximum error measure, when compared to a mean squared error measure.

\newpage

\renewcommand{\tabcolsep}{0.5em}
\renewcommand{\arraystretch}{1}
\begin{table}[h!]
\centering
\caption{Results for the four-dimensional Franke function: scaled RMSPE (black), scaled MAPE (blue), and time in minutes (red).}
  \begin{tabular}{|=l|+r|+r|+r|+r|+r|+r|+r|+r|}
    \hline
    \multicolumn{1}{|>{\columncolor{white}}c|}{} & \multicolumn{8}{c|}{Design Size}\\
    \arrayrulecolor{kugray5}
    \arrayrulecolor{black}
    \cline{2-9}
    \multicolumn{1}{|>{\columncolor{white}}c|}{} \rowstyle{\color{black}} & 9        & 41      & 129    & 321      & 681      & 1289      & 2241        & 3649 \\
    \hline
    \multirow{3}{*}{mlegp}                                   & 0.938 & 0.513  & 0.275 & 0.148   & 0.094   & 0.056     & 0.040       & 0.024 \\
                                    \rowstyle{\color{blue2}} & 0.988 & 0.456  & 0.382 & 0.204   & 0.228   & 0.119     & 0.133       & 0.098 \\
                                      \rowstyle{\color{red2}} & 0.015 & 0.034  & 0.254 & 2.759   & 20.828 & 139.565 & 791.228   & 4343.469 \\
    \hline
    \multirow{3}{*}{SGD}                                     & 0.719 & 0.395  & 0.173 & 0.095   & 0.054   & 0.027     & 0.020       & 0.009 \\
                                    \rowstyle{\color{blue2}} & 0.837 & 0.512  & 0.243 & 0.179   & 0.089   & 0.036     & 0.035       & 0.014 \\
                                      \rowstyle{\color{red2}} & 0.254 & 0.270  & 0.280 & 0.305   & 0.274   & 0.299     & 0.297       & 0.386 \\
    \hline
    \multirow{3}{*}{laGPsep}                               & 0.920 & 0.570  & 0.344 & 0.180   & 0.110   & 0.067     & 0.042       & 0.031 \\
                                    \rowstyle{\color{blue2}} & 0.933 & 0.569  & 0.498 & 0.301   & 0.322   & 0.140     & 0.095       & 0.093 \\
                                      \rowstyle{\color{red2}} & 0.028 & 1.030  & 2.699 & 5.805   & 11.822 & 17.150   & 17.141     & 17.179 \\
    \hline
    \multirow{3}{*}{tgp}                                        & 0.688 & 0.506  & 0.307 & 0.155   & 0.080   & 0.047     & 0.031       & 0.019 \\
                                    \rowstyle{\color{blue2}} & 0.654 & 0.595  & 0.393 & 0.233   & 0.185   & 0.094     & 0.070       & 0.051 \\
                                      \rowstyle{\color{red2}} & 0.487 & 1.263  & 4.752 & 20.561 & 76.153 & 285.270 & 1398.948 & 4790.997 \\
    \hline
    \multirow{3}{*}{SparseEm}                            & --        & 0.718  & --       & 0.450   & 0.436    & 0.411     & --              & 0.357 \\
                                    \rowstyle{\color{blue2}} & --         & 0.851 & --       & 0.470   & 0.514    & 0.435     & --              & 0.408 \\
                                      \rowstyle{\color{red2}} & --        & 2.543  & --       & 3.030   & 3.484    & 4.610    & --               & 11.482 \\
    \hline
  \end{tabular}
\label{tab:franke4dresults}
\end{table}

\renewcommand{\tabcolsep}{0.9em}
\renewcommand{\arraystretch}{1}
\begin{center}
\begin{table}[h!]
\center
\caption{APE results for the four-dimensional Franke function: scaled RMSPE (black), scaled MAPE (blue), and time in minutes (red).}
  \begin{tabular}{|=c|+r|+r|+r|+r|+r|+r|+r|+r|}
    \hline
    \multicolumn{1}{|>{\columncolor{white}}c|}{} & \multicolumn{6}{c|}{Design Size}\\
    \arrayrulecolor{kugray5}
    \arrayrulecolor{black}
    \cline{2-7}
    \multicolumn{1}{|>{\columncolor{white}}c|}{} & \textcolor{black}{300} & \textcolor{black}{503} & \textcolor{black}{999} & \textcolor{black}{1504} & \textcolor{black}{1998} & \textcolor{black}{3610} \\
    \hline
    \multirow{3}{*}{$n_0=100$}             & 0.209  & 0.157     & 0.125      & 0.085     & 0.070     & 0.050 \\
                       \rowstyle{\color{blue2}} & 0.348  & 0.205     & 0.239     & 0.170     & 0.170     & 0.125 \\
                        \rowstyle{\color{red2}} & 61.999 & 111.566 & 241.731 & 370.140 & 502.716 & 927.025 \\
    \hline
  \end{tabular}
\label{tab:franke4dresults2}
\end{table}
\end{center}

The SGD method outperforms the other methods for this test function, in terms of both predictive ability and computing time. Upon closer inspection of the sparse grid designs, many of the points appear to be located near the locations of the Gaussian peaks in Figure~\ref{fig:franke2d}. As a result, the designs capture the variability in the function well, and therefore result in high-quality fits. This is illustrative of an important feature of the SGD method: its performance depends heavily on the positions of the design points relative to the shape of the function.

APE is once again quite competitive with the others, although it does not match SGD in its predictive ability nor its speed. It is also slightly worse than laGPsep in both respects, although the computational complexity is still quite good. Overall, it does not have a clear advantage with this test function, as there is much more equal variability throughout the space.





\section{Discussion}
\label{sec:discussion}

Several conclusions can be made when comparing these methods across the two test functions. As expected, the standard GP fit using the \url{mlegp} package quickly becomes intractable, reaching a computing time of over three days for a fairly moderate problem.

The computing time for tgp is similar. Although tgp provides a very flexible model and introduces useful ideas for dealing with the large-dataset problem, the outer step of averaging over possible trees ultimately slows the method down significantly. This is fair, since fast computation was not a primary goal of tgp when it was first proposed.

SGD appears to perform very well in general. In some cases, it is competitive with mlegp in terms of predictive accuracy, while being quite fast. However, an important feature of this method is that its predictive performance depends on the positions of the design points relative to the shape of the function. For instance, for the corner peak function, the SGD did not contain any points close to the origin, where most of the variability lies (see Figure~\ref{fig:sgds}). As a result, its predictive ability for this function was much poorer.

One important point to note is that further tuning of parameters is required, in order to get optimal results with respect to each of the methods. In particular, SparseEm's performance may be sensitive to the choice of degree of sparsity within the correlation matrix, and decreasing this would likely further improve predictions (albeit at the expense of computing time).

There are several points to consider regarding the APE design method. The algorithm contains one main parameter: the initial design size $n_0$. This should ideally be tuned for optimal performance, according to the dimensionality and variability of the test function. In particular, the optimal value of $n_0$ remains an open problem.

There are also several aspects of the algorithm that may be adjusted in the future. The performance of the algorithm on functions such as the four-dimensional Franke function may be improved by implementing an exploratory aspect to the sequential search. Preliminary results have been obtained by choosing regions to split at random, with probabilities proportional to the CV prediction errors, instead of simply choosing the region with the maximum error.

In addition, preliminary results have also shown that the algorithm can be sped up by replacing the CV prediction errors with a simple measure of the variability of the responses in each region. In a preliminary study, this did not appear to have a highly negative effect on the overall predictive performance. Further work is required to study whether this is a viable choice in general.





\newpage

\bibliographystyle{apacite}
\bibliography{sample}

\begin{thebibliography}{}

\bibitem [\protect \citeauthoryear {%
Barthelmann%
, Novak%
\BCBL {}\ \BBA {} Ritter%
}{%
Barthelmann%
\ \protect \BOthers {.}}{%
{\protect \APACyear {2000}}%
}]{%
Barthelmannetal2000}
\APACinsertmetastar {%
Barthelmannetal2000}%
\begin{APACrefauthors}%
Barthelmann, V.%
, Novak, E.%
\BCBL {}\ \BBA {} Ritter, K.%
\end{APACrefauthors}%
\unskip\
\newblock
\APACrefYearMonthDay{2000}{}{}.
\newblock
{\BBOQ}\APACrefatitle {High dimensional polynomial interpolation on sparse
  grids} {High dimensional polynomial interpolation on sparse grids}.{\BBCQ}
\newblock
\APACjournalVolNumPages{Advances in Computational
  Mathematics}{12}{4}{273--288}.
\PrintBackRefs{\CurrentBib}

\bibitem [\protect \citeauthoryear {%
Chen%
}{%
Chen%
}{%
{\protect \APACyear {2018}}%
}]{%
Chen2018}
\APACinsertmetastar {%
Chen2018}%
\begin{APACrefauthors}%
Chen, H.%
\end{APACrefauthors}%
\unskip\
\newblock
\APACrefYear{2018}.
\unskip\
\newblock
\APACrefbtitle {Design and Analysis of Computer Experiments: Assessing and
  Advancing the State of the Art} {Design and analysis of computer experiments:
  Assessing and advancing the state of the art}\ \APACtypeAddressSchool
  {\BUPhD}{}{}.
\unskip\
\newblock
\APACaddressSchool {}{University of British Columbia}.
\PrintBackRefs{\CurrentBib}

\bibitem [\protect \citeauthoryear {%
Currin%
, Mitchell%
, Morris%
\BCBL {}\ \BBA {} Ylvisaker%
}{%
Currin%
\ \protect \BOthers {.}}{%
{\protect \APACyear {1991}}%
}]{%
Currinetal1991}
\APACinsertmetastar {%
Currinetal1991}%
\begin{APACrefauthors}%
Currin, C.%
, Mitchell, T.%
, Morris, M.%
\BCBL {}\ \BBA {} Ylvisaker, D.%
\end{APACrefauthors}%
\unskip\
\newblock
\APACrefYearMonthDay{1991}{}{}.
\newblock
{\BBOQ}\APACrefatitle {Bayesian prediction of deterministic functions, with
  applications to the design and analysis of computer experiments} {Bayesian
  prediction of deterministic functions, with applications to the design and
  analysis of computer experiments}.{\BBCQ}
\newblock
\APACjournalVolNumPages{Journal of the American Statistical
  Association}{86}{416}{953--963}.
\PrintBackRefs{\CurrentBib}

\bibitem [\protect \citeauthoryear {%
Dancik%
\ \BBA {} Dorman%
}{%
Dancik%
\ \BBA {} Dorman%
}{%
{\protect \APACyear {2008}}%
}]{%
mlegp}
\APACinsertmetastar {%
mlegp}%
\begin{APACrefauthors}%
Dancik, G\BPBI M.%
\BCBT {}\ \BBA {} Dorman, K\BPBI S.%
\end{APACrefauthors}%
\unskip\
\newblock
\APACrefYearMonthDay{2008}{}{}.
\newblock
{\BBOQ}\APACrefatitle {{mlegp}: Statistical analysis for computer models of
  biological systems using {R}} {{mlegp}: Statistical analysis for computer
  models of biological systems using {R}}.{\BBCQ}
\newblock
\APACjournalVolNumPages{Bioinformatics}{24}{17}{1966--1967}.
\PrintBackRefs{\CurrentBib}

\bibitem [\protect \citeauthoryear {%
Franke%
}{%
Franke%
}{%
{\protect \APACyear {1979}}%
}]{%
Franke1979}
\APACinsertmetastar {%
Franke1979}%
\begin{APACrefauthors}%
Franke, R.%
\end{APACrefauthors}%
\unskip\
\newblock
\APACrefYearMonthDay{1979}{}{}.
\newblock
\APACrefbtitle {A critical comparison of some methods for interpolation of
  scattered data} {A critical comparison of some methods for interpolation of
  scattered data}\ \APACbVolEdTR{}{\BTR{}}.
\newblock
\APACaddressInstitution{}{Naval Postgraduate School Monterey CA}.
\PrintBackRefs{\CurrentBib}

\bibitem [\protect \citeauthoryear {%
Gramacy%
}{%
Gramacy%
}{%
{\protect \APACyear {2007}}%
}]{%
tgp}
\APACinsertmetastar {%
tgp}%
\begin{APACrefauthors}%
Gramacy, R\BPBI B.%
\end{APACrefauthors}%
\unskip\
\newblock
\APACrefYearMonthDay{2007}{}{}.
\newblock
{\BBOQ}\APACrefatitle {{tgp}: An {R} package for {B}ayesian nonstationary,
  semiparametric nonlinear regression and design by treed {G}aussian process
  models} {{tgp}: An {R} package for {B}ayesian nonstationary, semiparametric
  nonlinear regression and design by treed {G}aussian process models}.{\BBCQ}
\newblock
\APACjournalVolNumPages{Journal of Statistical Software}{19}{9}{1--46}.
\newblock
\begin{APACrefDOI} \doi{10.18637/jss.v072.i01} \end{APACrefDOI}
\PrintBackRefs{\CurrentBib}

\bibitem [\protect \citeauthoryear {%
Gramacy%
}{%
Gramacy%
}{%
{\protect \APACyear {2016}}%
}]{%
laGP}
\APACinsertmetastar {%
laGP}%
\begin{APACrefauthors}%
Gramacy, R\BPBI B.%
\end{APACrefauthors}%
\unskip\
\newblock
\APACrefYearMonthDay{2016}{}{}.
\newblock
{\BBOQ}\APACrefatitle {{laGP}: Large-scale spatial modeling via local
  approximate {G}aussian processes in {R}} {{laGP}: Large-scale spatial
  modeling via local approximate {G}aussian processes in {R}}.{\BBCQ}
\newblock
\APACjournalVolNumPages{Journal of Statistical Software}{72}{1}{1--46}.
\newblock
\begin{APACrefDOI} \doi{10.18637/jss.v072.i01} \end{APACrefDOI}
\PrintBackRefs{\CurrentBib}

\bibitem [\protect \citeauthoryear {%
Gramacy%
\ \BBA {} Apley%
}{%
Gramacy%
\ \BBA {} Apley%
}{%
{\protect \APACyear {2015}}%
}]{%
GramacyApley2015}
\APACinsertmetastar {%
GramacyApley2015}%
\begin{APACrefauthors}%
Gramacy, R\BPBI B.%
\BCBT {}\ \BBA {} Apley, D\BPBI W.%
\end{APACrefauthors}%
\unskip\
\newblock
\APACrefYearMonthDay{2015}{}{}.
\newblock
{\BBOQ}\APACrefatitle {Local {G}aussian process approximation for large
  computer experiments} {Local {G}aussian process approximation for large
  computer experiments}.{\BBCQ}
\newblock
\APACjournalVolNumPages{Journal of Computational and Graphical
  Statistics}{24}{}{561--578}.
\PrintBackRefs{\CurrentBib}

\bibitem [\protect \citeauthoryear {%
Gramacy%
\ \BBA {} Lee%
}{%
Gramacy%
\ \BBA {} Lee%
}{%
{\protect \APACyear {2008}}%
}]{%
GramacyLee2008}
\APACinsertmetastar {%
GramacyLee2008}%
\begin{APACrefauthors}%
Gramacy, R\BPBI B.%
\BCBT {}\ \BBA {} Lee, H\BPBI K\BPBI H.%
\end{APACrefauthors}%
\unskip\
\newblock
\APACrefYearMonthDay{2008}{}{}.
\newblock
{\BBOQ}\APACrefatitle {Bayesian treed {G}aussian process models with an
  application to computer modeling} {Bayesian treed {G}aussian process models
  with an application to computer modeling}.{\BBCQ}
\newblock
\APACjournalVolNumPages{Journal of the American Statistical
  Association}{103}{}{1119--1130}.
\PrintBackRefs{\CurrentBib}

\bibitem [\protect \citeauthoryear {%
Jones%
, Schonlau%
\BCBL {}\ \BBA {} Welch%
}{%
Jones%
\ \protect \BOthers {.}}{%
{\protect \APACyear {1998}}%
}]{%
Jonesetal1998}
\APACinsertmetastar {%
Jonesetal1998}%
\begin{APACrefauthors}%
Jones, D\BPBI R.%
, Schonlau, M.%
\BCBL {}\ \BBA {} Welch, W\BPBI J.%
\end{APACrefauthors}%
\unskip\
\newblock
\APACrefYearMonthDay{1998}{}{}.
\newblock
{\BBOQ}\APACrefatitle {Efficient global optimization of expensive black-box
  functions} {Efficient global optimization of expensive black-box
  functions}.{\BBCQ}
\newblock
\APACjournalVolNumPages{Journal of Global Optimization}{13}{4}{455--492}.
\PrintBackRefs{\CurrentBib}

\bibitem [\protect \citeauthoryear {%
Kaufman%
}{%
Kaufman%
}{%
{\protect \APACyear {2010}}%
}]{%
SparseEm}
\APACinsertmetastar {%
SparseEm}%
\begin{APACrefauthors}%
Kaufman, C.%
\end{APACrefauthors}%
\unskip\
\newblock
\APACrefYearMonthDay{2010}{}{}.
\newblock
\APACrefbtitle {{SparseEm}: Statistical emulation using sparse correlation
  structure.} {{SparseEm}: Statistical emulation using sparse correlation
  structure.}
\newblock
\APAChowpublished {Retrieved Aug. 2, 2017, from
  \url{https://www.stat.berkeley.edu/~cgk/rcode/index.html}}.
\PrintBackRefs{\CurrentBib}

\bibitem [\protect \citeauthoryear {%
Kaufman%
, Bingham%
, Habib%
, Heitmann%
\BCBL {}\ \BBA {} Frieman%
}{%
Kaufman%
\ \protect \BOthers {.}}{%
{\protect \APACyear {2011}}%
}]{%
Kaufmanetal2011}
\APACinsertmetastar {%
Kaufmanetal2011}%
\begin{APACrefauthors}%
Kaufman, C.%
, Bingham, D.%
, Habib, S.%
, Heitmann, K.%
\BCBL {}\ \BBA {} Frieman, J\BPBI A.%
\end{APACrefauthors}%
\unskip\
\newblock
\APACrefYearMonthDay{2011}{}{}.
\newblock
{\BBOQ}\APACrefatitle {Efficient emulators of computer experiments using
  compactly supported correlation functions, with an application to cosmology}
  {Efficient emulators of computer experiments using compactly supported
  correlation functions, with an application to cosmology}.{\BBCQ}
\newblock
\APACjournalVolNumPages{Annals of Applied Statistics}{5}{}{2470--2492}.
\PrintBackRefs{\CurrentBib}

\bibitem [\protect \citeauthoryear {%
MATLAB%
}{%
MATLAB%
}{%
{\protect \APACyear {2017}}%
}]{%
MATLAB2017}
\APACinsertmetastar {%
MATLAB2017}%
\begin{APACrefauthors}%
MATLAB.%
\end{APACrefauthors}%
\unskip\
\newblock
\APACrefYear{2017}.
\newblock
\APACrefbtitle {Version R2017b} {Version r2017b}.
\newblock
\APACaddressPublisher{Natick, Massachusetts}{The MathWorks Inc.}
\PrintBackRefs{\CurrentBib}

\bibitem [\protect \citeauthoryear {%
McKay%
, Beckman%
\BCBL {}\ \BBA {} Conover%
}{%
McKay%
\ \protect \BOthers {.}}{%
{\protect \APACyear {1979}}%
}]{%
McKBecCon1979}
\APACinsertmetastar {%
McKBecCon1979}%
\begin{APACrefauthors}%
McKay, M\BPBI D.%
, Beckman, R\BPBI J.%
\BCBL {}\ \BBA {} Conover, W\BPBI J.%
\end{APACrefauthors}%
\unskip\
\newblock
\APACrefYearMonthDay{1979}{}{}.
\newblock
{\BBOQ}\APACrefatitle {A Comparison of Three Methods for Selecting Values of
  Input Variables in the Analysis of Output from a Computer Code} {A comparison
  of three methods for selecting values of input variables in the analysis of
  output from a computer code}.{\BBCQ}
\newblock
\APACjournalVolNumPages{Technometrics}{21}{2}{239--245}.
\PrintBackRefs{\CurrentBib}

\bibitem [\protect \citeauthoryear {%
O'Hagan%
}{%
O'Hagan%
}{%
{\protect \APACyear {1992}}%
}]{%
OHagan1992}
\APACinsertmetastar {%
OHagan1992}%
\begin{APACrefauthors}%
O'Hagan, A.%
\end{APACrefauthors}%
\unskip\
\newblock
\APACrefYearMonthDay{1992}{}{}.
\newblock
{\BBOQ}\APACrefatitle {Some {B}ayesian numerical analysis} {Some {B}ayesian
  numerical analysis}.{\BBCQ}
\newblock
\APACjournalVolNumPages{Bayesian Statistics}{4}{}{345--363}.
\PrintBackRefs{\CurrentBib}

\bibitem [\protect \citeauthoryear {%
Plumlee%
}{%
Plumlee%
}{%
{\protect \APACyear {2014}}%
{\protect \APACexlab {{\protect \BCnt {1}}}}}]{%
Plumlee2014}
\APACinsertmetastar {%
Plumlee2014}%
\begin{APACrefauthors}%
Plumlee, M.%
\end{APACrefauthors}%
\unskip\
\newblock
\APACrefYearMonthDay{2014{\protect \BCnt {1}}}{}{}.
\newblock
{\BBOQ}\APACrefatitle {Fast prediction of deterministic functions using sparse
  grid experimental designs} {Fast prediction of deterministic functions using
  sparse grid experimental designs}.{\BBCQ}
\newblock
\APACjournalVolNumPages{Journal of the American Statistical
  Association}{109}{}{1581--1591}.
\PrintBackRefs{\CurrentBib}

\bibitem [\protect \citeauthoryear {%
Plumlee%
}{%
Plumlee%
}{%
{\protect \APACyear {2014}}%
{\protect \APACexlab {{\protect \BCnt {2}}}}}]{%
SparseGridDesigns}
\APACinsertmetastar {%
SparseGridDesigns}%
\begin{APACrefauthors}%
Plumlee, M.%
\end{APACrefauthors}%
\unskip\
\newblock
\APACrefYearMonthDay{2014{\protect \BCnt {2}}}{}{}.
\newblock
\APACrefbtitle {{Sparse Grid Designs}: {MATLAB} package, version 1.5.0.0.}
  {{Sparse Grid Designs}: {MATLAB} package, version 1.5.0.0.}
\newblock
\APAChowpublished {Retrieved Oct. 2016, from
  \url{https://www.mathworks.com/matlabcentral/fileexchange/45668-sparse-grid-designs}}.
\PrintBackRefs{\CurrentBib}

\bibitem [\protect \citeauthoryear {%
{R Core Team}%
}{%
{R Core Team}%
}{%
{\protect \APACyear {2017}}%
}]{%
RCoreTeam2017}
\APACinsertmetastar {%
RCoreTeam2017}%
\begin{APACrefauthors}%
{R Core Team}.%
\end{APACrefauthors}%
\unskip\
\newblock
\APACrefYearMonthDay{2017}{}{}.
\newblock
{\BBOQ}\APACrefatitle {R: A Language and Environment for Statistical Computing}
  {R: A language and environment for statistical computing}{\BBCQ}\
  [\bibcomputersoftwaremanual].
\newblock
\APACaddressPublisher{Vienna, Austria}{}.
\newblock
\begin{APACrefURL} \url{https://www.R-project.org/} \end{APACrefURL}
\PrintBackRefs{\CurrentBib}

\bibitem [\protect \citeauthoryear {%
Sacks%
, Welch%
, Mitchell%
\BCBL {}\ \BBA {} Wynn%
}{%
Sacks%
\ \protect \BOthers {.}}{%
{\protect \APACyear {1989}}%
}]{%
Sacksetal1989}
\APACinsertmetastar {%
Sacksetal1989}%
\begin{APACrefauthors}%
Sacks, J.%
, Welch, W\BPBI J.%
, Mitchell, T\BPBI J.%
\BCBL {}\ \BBA {} Wynn, H\BPBI P.%
\end{APACrefauthors}%
\unskip\
\newblock
\APACrefYearMonthDay{1989}{}{}.
\newblock
{\BBOQ}\APACrefatitle {Design and analysis of computer experiments} {Design and
  analysis of computer experiments}.{\BBCQ}
\newblock
\APACjournalVolNumPages{Statistical Science}{4}{}{409--423}.
\PrintBackRefs{\CurrentBib}

\bibitem [\protect \citeauthoryear {%
Stein%
, Chi%
\BCBL {}\ \BBA {} Welty%
}{%
Stein%
\ \protect \BOthers {.}}{%
{\protect \APACyear {2004}}%
}]{%
Steinetal2004}
\APACinsertmetastar {%
Steinetal2004}%
\begin{APACrefauthors}%
Stein, M\BPBI L.%
, Chi, Z.%
\BCBL {}\ \BBA {} Welty, L\BPBI J.%
\end{APACrefauthors}%
\unskip\
\newblock
\APACrefYearMonthDay{2004}{}{}.
\newblock
{\BBOQ}\APACrefatitle {Approximating likelihoods for large spatial data sets}
  {Approximating likelihoods for large spatial data sets}.{\BBCQ}
\newblock
\APACjournalVolNumPages{Journal of the Royal Statistical Society: Series B
  (Statistical Methodology)}{66}{2}{275--296}.
\PrintBackRefs{\CurrentBib}

\bibitem [\protect \citeauthoryear {%
Surjanovic%
\ \BBA {} Bingham%
}{%
Surjanovic%
\ \BBA {} Bingham%
}{%
{\protect \APACyear {2013}}%
}]{%
SurjanovicBingham2013}
\APACinsertmetastar {%
SurjanovicBingham2013}%
\begin{APACrefauthors}%
Surjanovic, S.%
\BCBT {}\ \BBA {} Bingham, D.%
\end{APACrefauthors}%
\unskip\
\newblock
\APACrefYearMonthDay{2013}{}{}.
\newblock
\APACrefbtitle {Virtual Library of Simulation Experiments: Test Functions and
  Datasets.} {Virtual library of simulation experiments: Test functions and
  datasets.}
\newblock
\APAChowpublished {Retrieved from \url{http://www.sfu.ca/~ssurjano}}.
\PrintBackRefs{\CurrentBib}

\end{thebibliography}

\end{document}